\documentclass[12pt,a4paper,fleqn]{article}

\usepackage[textheight=23cm,textwidth=16cm]{geometry}
\usepackage{amsmath}
\usepackage{graphicx}
\usepackage[american]{babel}
\usepackage{here}
\usepackage[articletitle=true]{achemso}

\begin{document}

\begin{center}

{\LARGE\bf
 Heuristics and Uncertainty Quantification in Rational and Inverse Compound and Catalyst Design
}

\vspace{1cm}

{\large
Thomas Weymuth\footnote{e-mail: thomas.weymuth@phys.chem.ethz.ch; ORCID: 0000-0001-7102-7022} and
Markus Reiher\footnote{Corresponding author; e-mail: markus.reiher@phys.chem.ethz.ch; ORCID: 0000-0002-9508-1565}
}\\[4ex]

Laboratory of Physical Chemistry and National Centre of Competence in Research (NCCR) Catalysis, ETH Zurich, Vladimir-Prelog-Weg 2, 8093 Zurich, Switzerland

March 17, 2022

\vspace{.5cm}

\textbf{Abstract}
\end{center}
\vspace*{-.41cm}
{
The goal of inverse (quantum) approaches is to devise methods and approaches capable of efficiently searching chemical space in such a way that the design of novel materials and compounds with specific properties is as direct and efficient as possible.
Here, we review the current state of the field with a focus on the most recent developments. We discuss the importance of heuristic rules and human intuition for rational compound design. Moreover, we elaborate on options for reliable uncertainty quantification for computational results, which is crucial for a truly predictive application of any \textit{in silico} method.
}


\vspace{.7cm}

\textbf{Key Points}
\begin{itemize}
\item Inverse quantum mechanical approaches enable one to search for previously unknown molecules with predefined properties.
\item Despite all advances, chemical space is too large to be exhaustively searched; the search for new molecules must be restricted to a suitably defined subspace. For the definition of such a subspace, heuristics and intuition are of great importance.
\item The quantification of the uncertainty of calculated data is crucial for a successful application of inverse methods.
\item This article reviews the current state of inverse (quantum) approaches, emphasizing the developments of the past ten years.
\end{itemize}

\newpage

\section{Introduction}
\label{sec:introduction}

A wide range of quantum mechanical methods is available that---together with the currently
available computational resources---allow for routine calculations on molecules with hundreds of atoms\cite{Dykstra2005}.
As a result, extensive first-principles molecular dynamics simulations have become standard applications\cite{Meuwly2019}
and automated explorations of vast chemical reaction networks are feasible\cite{Dewyer2018, Vazquez2018, Simm2019, Unsleber2020, Shannon2021, Maeda2021, Steiner2021}.

Underlying all these calculations is the solution of the electronic Schr\"odinger equation,
\begin{equation}
  \hat{H}_{\rm el}\Psi_{\rm el} = E_{\rm el}\Psi_{\rm el}~,
\end{equation}
which results in the Born--Oppenheimer approximation\cite{Born1927, Tully2000}
that separates electronic and nuclear motion in such a way that the electronic motion produces a potential energy surface, $E_{\rm el}$, on which the nuclei move (either described by wave packets or by Newtonian trajectories when treated as classical particles). 
$\Psi_{\rm el}$ is the electronic wave function describing the electronic system under consideration and $\hat{H}_{\rm el}$ is the corresponding electronic Hamiltonian, whose 
nonrelativistic form is exact and, in Hartree atomic units, reads 
\begin{equation}
\label{eq:h}
  \hat{H} =             -\frac{1}{2}\sum_i \Delta_i 
            -\sum_{i,I} \frac{Z_I}{r_{iI}}
            +\sum_{i<j} \frac{1}{r_{ij}}
            +\sum_{I<J} \frac{Z_I Z_J}{r_{IJ}}~,
\end{equation}
where the subscripts $I$ and $J$ denote atomic nuclei, while $i$ and $j$ denote electrons. $M_I$ is the mass of
nucleus $I$, while $Z_I$ is its charge number. $r_{ij}$ is the distance between the particles $i$ and $j$ and the
Laplacian is defined as
\begin{equation}
    \Delta_i = \frac{\partial^2}{\partial x_i^2} +
               \frac{\partial^2}{\partial y_i^2} +
               \frac{\partial^2}{\partial z_i^2}~.
\end{equation}
The first sum of terms in
Eq.~(\ref{eq:h}) denotes the kinetic energy of the electrons, while the remaining three classes of terms denote the
electrostatic Coulomb interaction between electrons and nuclei.

The electronic Hamiltonian contains, as parameters, the charges and positions of all atomic nuclei, as well as the total number of electrons.
In other words, one needs to specify a molecular structure and total charge before the electronic Schr\"odinger equation can be solved. Only then, properties such as electronic energies and their derivatives, which are identified as molecular properties\cite{Norman2018},
can be calculated.
However, in many practical situations such as the design of new materials, the problem is not to compute the properties for a given
structure (\textit{i.e.}, scaffold of nuclei). Rather, a certain property (or set of properties) is desired and the structure of a compound or material featuring this property is sought for. Hence, when considering the solution of the Schr\"odinger equation as the forward way, the typical atomistic design problem is an inverse problem that requires reversing the numerical solution process in some way.
The idea of inverse quantum chemistry and inverse quantum materials science is, conceptually speaking, to invert the electronic Schr\"odinger equation in order to go from property
to structure rather than from structure to property. Whereas the latter approach clearly points to high-throughput screening in search of desired properties and function, the former aims at a directed approach that allows one to point to structural features that enable certain properties and functions.  

A direct inversion of the electronic Schr\"odinger equation or its descendants will not be possible in general\cite{Weymuth2014a, Zunger2018} (however, see \textit{e.g.},
Ref.~\citenum{Sabatier2000} for an overview of inverse mathematical methods developed). This is the reason why current inverse approaches usually feature protocols that
solve the Schr\"odinger equation in the forward direction for a vast collection of structures, until
substances with a desired property are found. The biggest challenge for this approach is the huge size of chemical space: For instance, the GDB-17 database contains over
166 billion molecules\cite{Ruddigkeit2012, Reymond2015}; the number of molecules containing up to thirty carbon, nitrogen, oxygen, and sulfur atoms has
been estimated to be larger than $10^{60}$\cite{Bohacek1996}, while the number of proteins that could exist is about $10^{390}$\cite{Dobson2004}
(for a size of 300 amino acid residues per protein).

Therefore, ways need to be devised to improve this process. One possibility is to employ advanced sampling algorithms, minimizing the
number of candidate structures to be probed (ideally with fast 
approximations for the solution of the Schr\"odinger equation). The most promising recent approach to achieve acceleration of the screening is by  machine learning, which basically
creates a surrogate function to approximate a solution of the Schr\"odinger equation that is very fast to evaluate once trained.

One aspect that is largely ignored in inverse
quantum chemistry is uncertainty quantification. As every quantum chemical calculation has a certain inherent error, it is
important to attach an error bar to every quantum chemical result in order to be able to judge its reliability. Unfortunately, estimating
the uncertainty of quantum chemical calculations is far from trivial. Still, progress has been made.

Here, we review the most recent developments in the field of inverse quantum chemistry, with a focus on the
past decade (for further discussions see Refs.~\citenum{Weymuth2014a, Zunger2018, Freeze2019}).
In particular, we will
discuss heuristics and uncertainty quantification for inverse quantum chemical algorithms.

\section{Quantitative Structure--Property Relationships}

The idea that a certain property depends in a systematic way on the underlying molecular structure is a very old one. Already in 1863, Cros described a relation between the toxicity of primary aliphatic alcohols and their solubility in water\cite{Cros1863}. Very well known is Hammett's work relating reaction rates and equilibrium constants for reactions involving substituted benzene derivatives\cite{Hammett1935, Hammett1937}. However, the starting point of modern quantitative structure--property relationships (QSPR) and the closely related quantitative structure--activity relationships (QSAR) studies is widely regarded as being the work by Hansch \textit{et al.}~in the early 1960s\cite{Hansch1962, Hansch1963, Hansch1964}. These authors built on the work by Hammett and showed how it can be used to obtain mathematical relationships between closely related molecular structures and a range of properties such as toxicity.

The basic goal of any QSPR study is to empirically find some mathematical function relating a specific molecular property (or set of such properties) to molecular structure\cite{Katritzky1995, Karelson1996, Katritzky2010, Le2012, Berhanu2012, Pirhadi2015, Polishchuk2017, Muratov2020}. In this mathematical function, the molecular structure may be described in a wide variety of ways. For example, simple descriptors such as bond lengths and bond angles or the number of ring systems might be used. However, also more advanced concepts such as the Tanimoto similarity\cite{Rogers1960} are used\cite{Nikolova2003}. It is, however, often not clear \textit{a priori} which structural descriptors are best suited, neither is the mathematical functional dependence obvious. Therefore, carrying out a successful QSPR analysis is far from trivial and cannot be automated in a simple way.

Two main classes of QSPR studies may be distinguished\cite{Fujita2016}. On the one hand, there are studies relying on a comparatively small data set of molecules. Often, conceptually simple mathematical relationships such as linear functions are employed. 
Such comparatively simple QSPR relationships 
have the great advantage of straightforward interpretation. Moreover, they can provide heuristic models. One example is the Sabatier principle, which relates the efficacy of a heterogeneous catalyst to a property like the heat of adsorption of the substrate on the catalyst surface\cite{Rothenberg2008}.
However, one of the main drawbacks of this class of studies is that the QSPR derived are often only applicable to molecules which closely resemble each other. Hence, they are not generally applicable throughout the entire chemical space.

On the other hand, more and more studies involve very large molecular data sets. This includes data-mining approaches relying on large databases such as the Cambridge Structure Database\cite{Groom2016}. An example of this approach is the search for new energy materials as reviewed recently by Zhand and others\cite{Zhang2019}. Large data sets facilitate the training of sophisticated machine-learning (ML) models such as neural networks. However, these
models are not as straightforward to interpret as, for example, a simple linear function, but they are usually much more generally applicable.

Once a QSPR model has been set up, finding the structural descriptor(s) leading to a certain desired property value is often straightforward, even though it can also be challenging for complex models (see, \textit{e.g.}, Ref.~\citenum{Miyao2016}). In either case, it is usually a highly nontrivial task to identify actual molecular structures corresponding to these optimal descriptor values. 

The field of inverse QSPR/QSAR addresses this problem\cite{Zheng1998, Cho1998, Gantzer2020}. The basic concept of all inverse QSPR approaches is to search a specific fraction of chemical space in order to find the most favorable structure as predicted by a given QSPR model. The chemical subspace searched can be predefined in terms of a fixed set of molecules (such as the GDB-17 database\cite{Ruddigkeit2012, Reymond2015}) or it can be dynamically constructed, \textit{e.g.}, by combining predefined molecular fragments to yield new molecules. The search itself can be carried out in a
brute-force manner or by advanced sampling approaches such as evolutionary algorithms (see also section~\ref{sec:sampling}).

\section{High-Throughput Virtual Screening}

High-throughput virtual screening
can be a key component of inverse quantum chemistry and rational compound design\cite{Shoichet2004, Pyzer-Knapp2015}. 
Early high-throughput screening techniques have already been employed in the pharmaceutical industry starting in the 1990s\cite{Attene-Ramos2014, Bevan1995}. When computational resources became widely available, the concept of high-throughput screening could be adapted to a computational approach in which a given database of compounds is screened by theoretical rather than experimental methods, \textit{i.e.}, for every single substance the property of interest is calculated, not measured.

One might argue that such a screening approach is not truly inverse, since it is essentially just a series of ``forward'' calculations in which the Schr\"odinger equation is solved with conventional means. However, it is important to realize that currently all practical inverse approaches involve at their core such forward calculations (\textit{cf.}, the introduction). In this respect, high-throughput virtual screening can be regarded as an inverse approach. While brute-force screening of a predefined database might not always be the most efficient way of finding a given target compound, the data created is extremely helpful in the creation of machine-learning models (see below).

One example of high-throughput virtual screening is the Harvard Clean
Energy Project (CEP)\cite{Hachmann2011, Hachmann2014}. The goal of the project was the identification of new materials for organic solar cells. Through a combinatorial molecule generator with 26 building blocks, the CEP created a library of about 10'000'000 candidate molecules. These molecules were subsequently all tested for their suitability as photovoltaic materials. For this screening, the CEP relied in large parts on the World Community Grid, a distributed volunteer computing platform organized by IBM\cite{Clery2005}. With this massive computing power available, the CEP screened about 20'000 compounds per day\cite{Hachmann2011}. In total, about 10 million compounds were screened; about 1000 compounds were found to be promising for photovoltaic applications\cite{Pollice2021}.

Related to high-throughput virtual screening are  data-mining approaches such as the clustering--ranking--modeling (CRM) methodology\cite{Meredig2014}. In CRM, a (potentially huge) database of compounds is evaluated with respect to a given property, as in traditional high-throughput virtual screening approaches. However, if the property of interest could only be obtained in a computationally expensive calculation, it had not been calculated directly, but was approximated by another descriptor whose evaluation was much faster. Of course, a correlation between the two descriptors needs to be validated beforehand---this essentially amounts to establishing a QSPR (see above). It is accomplished by setting up a training database containing a range of possible descriptors, clustering all compounds and then searching for correlations between descriptors for every cluster. Besides the reduced computational cost, CRM can also offer unexpected insights by revealing non-obvious correlations.

\section{Advanced Sampling Methods}
\label{sec:sampling}

High-throughput virtual screening, although simple to implement and to carry out, is often a very time-consuming process. In addition, because of the brute-force approach taken, many molecules are considered which have properties that may be unsuitable for the given purpose. Therefore, a wide range of advanced sampling methods has been devised, all with the goal to reduce the chemical space or, in other words, the number of candidate molecules that have to be considered before a compound with the desired property is found.

A suitable algorithm for global optimization problems in general and rational compound design in particular is simulated annealing\cite{Kirkpatrick1983, Kirkpatrick1984}. The name of this algorithm derives from the process of annealing in materials science, in which a heat treatment with controlled temperature is used to increase the ductility of a material. Likewise, in simulated annealing, there is a (virtual) temperature parameter which is set to a comparatively high value at the beginning and then continuously lowered over the course of the optimization. While searching, the algorithm evaluates new points according to a fitness function and accepts them according to a Metropolis criterion---the higher the temperature, the higher the probability that a point with a comparatively bad fitness is accepted. This allows the algorithm to overcome barriers. Towards the end of the algorithm, when the temperature tends toward its final target value, the algorithm settles for the solution with the best fitness.

Simulated annealing was key, for example, for the work by Sch\"on and Jansen toward the facilitation of synthesis planning in solid state chemistry\cite{Schon1996, Jansen2002}, where random unit cells were created and then the positions and composition of the atoms as well as the unit cell vectors were optimized.

A similar approach has been presented by Franceschetti and Zunger in 1999\cite{Franceschetti1999} who started from a predefined (random) unit cell and then optimized its composition by simulated annealing such that the material
exhibits a desired property. A few years later, Dudiy and Zunger adapted this approach to a genetic algorithm for the optimization step\cite{Dudiy2006}, which usually leads to faster convergence. This methodology has since then been applied repeatedly in the inverse design of new materials (see, \textit{e.g.}, Ref.~\citenum{Avezac2012}). 
In inverse molecular design, genetic algorithms have been used at least since 1993\cite{Venkatasubramanian1994}. Today, many different frameworks rely on evolutionary algorithms for rational compound design, but their differences are often only subtle.

Genetic algorithms are a subclass of evolutionary algorithms. These are inspired by biological evolution and processes such as mutation and recombination.
For a recent review on the application of evolutionary algorithms in materials design, we refer the reader to Ref.~\citenum{Le2016}.
In evolutionary algorithms, candidate solutions 
play the role of individuals in a population; over the course of the algorithm, new individuals are created by combining existing candidate solutions and by introducing (small) random changes into existing solutions. Individuals with a low fitness are continuously eliminated from the population. Therefore, the population is gradually steered towards the best solution.
Evolutionary algorithms are very often applied with great success in complex optimization problems. 

Since the development of rational design frameworks is still in its infancy and mostly prototypical work has been accomplished, we list some advances in the following.

One such framework is called ``NovoFLAP'', presented in 2010 by Damewood, Jr.~and others\cite{Damewood2010}. Starting from a set of predefined structures, new candidate molecules are generated with a fragment-based approach. These are then evaluated according to a pharmacological scoring function developed at the company AstraZeneca. The primary aim of NovoFLAP is to create new medicinally relevant structural motifs, \textit{i.e.}, for finding new leads. 

Developed in 2013, ``Algorithm for Chemical Space Exploration with Stochastic Search'' (ACSESS)\cite{Virshup2013, Rupakheti2015} is a framework for rational compound design that starts from a given set of compounds. ACSESS then automatically creates new candidate molecules following a set of rules called ``chemical mutations''. Molecules which are not considered synthesizable or do not meet other predefined constraints (such as, for example, having a certain substructure), are eliminated. Then, a maximally diverse set of molecules is selected and the properties of interest are calculated for these molecules to probe whether any of these molecules exhibits the desired properties to a sufficient degree. These steps are then iterated until at least one target molecule has been found. ACSESS has been developed with a special focus on the rigorous exploration of chemical subspaces, even when these are very large.

Another framework is ``Poor Man's Materials Optimization'' (PooMa)\cite{Springborg2017, Huwig2017} which is conceptually very similar to ACSESS in that it automatically constructs candidate molecules and searches among them for target compounds using a genetic algorithm. However, PooMa is designed with simplicity and efficacy in mind, and hence, a standard workstation or laptop shall be sufficient to run the program\cite{Huwig2017}.

A fourth framework is the ``Multiobjective Evolutionary Graph
Algorithm'' (MEGA)\cite{Nicolaou2009}, where the core concepts are again identical to the aforementioned programs, but the focus of MEGA is on the simultaneous optimization of several target properties at once. A key conceptual feature of MEGA is that the search space is represented as a graph.

Very recently, Jensen and coworkers presented the program  DENOPTIM\cite{Foscato2019}, which is a general-purpose package for \textit{de novo} design of both organic and inorganic compounds. While it relies on an evolutionary algorithm for an efficient search, it also allows for the brute-force screening of an entire library.

Lameijer \textit{et al.}~proposed the ``Molecule Evoluator'', a program featuring interactive evolution\cite{Lameijer2006}, where the operator assumes the role of the fitness function directing the evolutionary algorithm. This allows her or him to incorporate human chemical intuition and creativity into the design process.

Kawai \textit{et al.}~presented an approach which attempts to find new molecules which are structurally similar to a reference molecule\cite{Kawai2014}. This similarity is used to constrain the otherwise vast chemical space to be searched. At the same time, one can expect with reasonable confidence that the proposed molecules are chemically feasible (provided the reference molecule can be synthesized). Note that a similar approach, also attempting to find new molecules which are similar to a reference molecule, has already been presented in 2006 by Fechner and Schneider\cite{Fechner2006, Fechner2007}.

Another algorithm is the best-first search method\cite{Pearl1987}. This algorithm belongs to the class of so-called branch-and-bound techniques; the search space is represented as a graph, and the most promising node is evaluated first. The graph is traversed until a node which satisfies all desired properties is found. De Proft and coworkers have proposed a framework relying on best-first search as the optimization algorithm in inverse design\cite{DeVleeschouwer2012, DeVleeschouwer2016}.

In 2014, Weymuth and Reiher proposed ``Gradient-driven Molecule Construction'' (GdMC)\cite{Weymuth2014b} which assumes that a local property can be represented by a specific local structural feature. The basic idea is then to start with this local feature and consider it as an incomplete molecular scaffold that needs to be stabilized by a surrounding molecular structure into which it is embedded in a chemically meaningful way. This environment may be abstracted as a ``jacket potential'' that stabilizes the local scaffold (through vanishing nuclear gradients) and can be seen as a target for the design of a proper (\textit{i.e.}, stable) molecular structure\cite{Weymuth2013}. Hence, in a second step, the jacket potential has to be converted  to a concrete chemical structure, for which various options exist. For example, it can be represented numerically on a grid (this is especially useful in the framework of density functional theory, where such a grid is usually present in practical calculations) or by point charges. Alternatively, one can directly add additional nuclei and electrons or even entire structural fragments to the local structural element, thereby avoiding the abstract representation of the jacket potential altogether. Reiher and coworkers have demonstrated an automatized procedure for a shell-wise buildup of such a ligand sphere, and have shown how it can successfully reproduce a ruthenium catalyst known to bind and activate CO\textsubscript{2} and propose further potentially suitable catalysts\cite{Krausbeck2017}.

A common challenge for all the above approaches is the creation of new, chemically reasonable candidate molecules. It is not so much of a problem when relying on a library of predefined structural fragments (although even in this case, certain geometric constraints have to be fulfilled), but this approach has the disadvantage that the chemical subspace which can be searched is intrinsically limited by the kind of fragments available. An alternative to this is to represent chemical structures as so-called SMILES strings; this is often done in machine learning-based approaches (see, \textit{e.g.}, Ref.~\citenum{Ikebata2017} and section~\ref{sec:ml}). A third common approach is to use an abstract representation during the optimization---such as the jacket potential in GdMC---and only convert this representation to a real chemical structure once the optimization is finished. This approach has the additional advantage that it renders the chemical subspace in a truly continuous manner, which is beneficial for many optimization routines. An example of such an approach is the globally optimal catalytic field method presented by Dittner and Hartke\cite{Dittner2018}. Their idea is to increase the efficiency of a given catalyst by embedding it into an abstract environment. Note that this approach also allows for the \textit{de novo} construction of new catalysts---in this case, the abstract environment plays the role of the entire catalyst, rather than just a part of it. 
After an optimal environment has been found, the task is to convert it to an actual chemical structure. In their first study, Dittner and Hartke refrained to tackle this second step and simplified the representation and optimization of the abstract environment insofar as they used simple point charges placed on a van der Waals surface around the catalyst\cite{Dittner2018}. They were able to find point charge collections which had a significant catalytic effect on the prototype Menshutkin reaction. Dittner and Hartke improved their method and were able to successfully apply it to a prototypical Diels--Alder reaction\cite{Dittner2020}. Very recently, Behrens and Hartke addressed the problem of chemical representability of the abstract environment\cite{Behrens2021}. They decided that the easiest way for this is to abandon an abstract representation altogether, instead relying on a library of predefined fragments.

In 2006, Wang \textit{et al.}~presented a possible solution to the problem of deriving a chemical structure from an abstract representation with an approach called ``linear combination of atomic potentials'' (LCAP)\cite{Wang2006}. In essence, this approach expands the abstract representation into a set of given atomic potentials for which the chemical representation is known (although called ``atomic potentials'', they are not limited to represent individual atoms, but could also represent more complex fragments made up of several atoms). Then, only the number, kind, spatial positions and the expansion coefficients of these atomic potentials must be optimized. In practical applications, however, the spatial positions at which atomic potentials might be placed are predefined, as are the kinds of potentials available. The LCAP methodology has been continuously improved during the last years. For example, very recently, Shiraogawa and Ehara took advantage of it in the design of photofunctional molecular aggregates\cite{Shiraogawa2020}.

A third approach to render chemical space continuous has been proposed by von Lilienfeld \textit{et al}.~in 2005\cite{vonLilienfeld2005}. Starting in the field of conceptual density functional theory, these authors derived an expression for the derivative of the total electronic energy with respect to the nuclear charge distribution. They proposed to call it ``alchemical potential'', since it is a measure for the tendency of a molecule to ``transmutate'', or change, a given atomic nucleus to a different one. In this spirit, one can define an alchemical transformation as any physical or fictitious process  that connects two points in chemical space. We note that the term ``computational alchemy'' is not new\cite{Straatsma1992}; in fact, the basic concept of alchemical transformations goes back to the well-known thermodynamic integration\cite{Kirkwood1935}. Although they are not always trivial to evaluate (see, \textit{e.g.}, Ref.~\citenum{Munoz2017}), the availability of analytical property derivatives for these potentials can make searching through this space very efficient\cite{Chang2014, toBaben2016, Saravanan2017, Domenichini2020}.

Irrespective of whether the search space is discrete or continuous, every optimization algorithm needs a starting point, \textit{i.e.}, at least one chemical structure from which it begins to search for optimized structures. The choice of this starting point is of crucial importance: if it is in some sense close to the desired optimum, the task of finding this optimum is greatly simplified---in the best case, even a straightforward gradient descent algorithm might then be able to find the optimum. However, it is often not obvious how to choose the starting structures. Usually, human chemical intuition and heuristics are required to come up with a suitable set of initial molecules. For example, a typical procedure would be to analyze the results of previous screening studies and select a number of molecules the properties of which are closest to the desired target properties. However, such an approach does not guarantee that a molecule with the desired properties is actually found.

\section{Machine-Learning Models}
\label{sec:ml}

Both high-throughput screening and the application of any of the sampling methods presented above can yield vast amounts of data, which lend themselves to construct machine-learning models. Accordingly, ML-based approaches now form a cornerstone in inverse quantum chemistry\cite{Sanchez-lengeling2018, Mater2019, vonLilienfeld2020, Pollice2021, Jena2021, Jena2021a, Nandy2021, Janet2021, Huang2021}. They offer a significant speedup compared to the methods discussed above\cite{Teunissen2019}.

In general, ML models are intrinsically interpolative in nature. Therefore, the accuracy of their predictions can quickly become questionable outside their range of parametrization. This is of particular importance in the field of inverse design, because new candidate molecules could be in a region of chemical space which was not considered for the training of the ML model. In 2016, Patra \textit{et al.}~proposed an elegant solution to this problem\cite{Patra2017}. The key idea is that, once the candidate molecules proposed by an ML model have been evaluated by a quantum chemical calculation or an experiment, the results of this evaluation are fed back and the ML model is retrained. In this way, the underlying ML model is constantly retrained and improved in a rolling fashion.

For an overview on ML approaches in the context of rational and inverse design, we may mention a few examples: Mannodi-Kanakkithodi \textit{et al.}~combined an ML model based on kernel-ridge regression with a genetic algorithm to search new polymer
dielectrics\cite{Mannodi-Kanakkithodi2016}. Hautier and others employed an ML model to search for new ternary oxides, identifying 209 previously unknown compounds\cite{Hautier2010}. Yoshida and coworkers presented a Bayesian model and demonstrated its application by designing organic molecules with a predefined HOMO--LUMO gap\cite{Ikebata2017}. Raccuglia \textit{et al.}~explicitly incorporated data from unsuccessful experiments to build a support vector machine
model guiding future syntheses\cite{Raccuglia2016}. Zare and coworkers presented a deep reinforcement learning framework for the optimization of reaction conditions\cite{Zhou2017}. Kulik and coworkers demonstrated how machine learning can be employed in complex multiobjective optimization problems\cite{Janet2020}.

\section{Uncertainty Quantification}

Every (quantum) calculation has a certain inherent error. Typically, faster methods are less accurate, which means that their error is larger compared to methods which are more computer time consuming because they aim at approximating an accurate result in a systematic way. However, even comparatively small errors can have a large effect. For example, when activation energies are used to calculate reaction rate constants in an attempt to provide data on kinetic stability or synthetic accessibility, small uncertainties in these activation energies translate into rather large uncertainties of the rate constants due to the exponential form of the Arrhenius equation\cite{Proppe2016, Proppe2019}.

Quantifying this error, \textit{i.e.}, the uncertainty of a certain quantum chemical calculation is far from being trivial. The primary means of establishing the accuracy of any quantum chemical method has been benchmarking, \textit{i.e.}, comparing results obtained with some method to those of a reference method or experiment for atomistic structures that are assumed to be relevant for a case or target at hand. However, the error of a given method is not constant, or homoscedastic, across chemical space, but can fluctuate considerably from one molecule to another, \textit{i.e.}, it behaves heteroscedastically. This, in turn, implies that results of benchmark studies are in general not transferable to regions of chemical space different to the ones covered in this benchmark.

Therefore, having a reliable error bar for the specific results that one is interested in would be desirable. This is particularly true in the field of inverse quantum chemistry, where one can easily come across exotic molecules which have never been covered by any benchmark set---hence, one can be completely in the dark as to the accuracy of the method employed. Moreover, if inverse design problems involve searching through a large chemical subspace, and therefore, a huge number of quantum chemical calculations must be carried out,
then it will be of utmost importance to be able to rely on very fast computational schemes so that the question of system-specific uncertainty becomes even more urgent\cite{Reiher2021}.
If a reliable (and sufficiently small) error bar were available in such cases, one could employ fast methods with confidence, and recalculate the specific cases in which the estimated uncertainty is above a given threshold.

Besides benchmarking, a different way of estimating the uncertainty associated with a given result is to use scientific judgment. In fact, this is referred to as ``type B evaluation of standard uncertainty'' in the ``Guide to the expression of uncertainty in measurement'' published by the joint technical committee of the International Organization for Standardization and the International Electrotechnical Commission\cite{iso_iec_guide_98_3}. In certain circumstances, this is the only viable way to come up with an uncertainty. However, concerning quantum chemical results, this way of manually estimating the uncertainty is extremely complex and error-prone. Therefore, it is usually not done.

The main problem of estimating the error associated with a certain computational result is that it is entirely systematic, which means that the standard way to obtain the uncertainty of a laboratory measurement, namely to repeatedly run the computation followed by obtaining the standard deviation of the results, is meaningless because one will always obtain the same result to high precision. However, the ``Guide to the expression of uncertainty in measurement''\cite{iso_iec_guide_98_3} mentions a way to quantify the uncertainty arising from systematic errors (often called bias). Based on this, Irikura \textit{et al.}~proposed a way to correct this bias of so-called virtual measurements and to estimate the uncertainty of the final, corrected result\cite{Irikura2004}. In essence, their proposal is to apply a correction to the computational result. This can be done, for example, by obtaining the properties of interest for molecules which are similar to the molecules of interest and for which high-quality data (\textit{e.g.}, from experiments) is available. Then, a mathematical expression for a correction can be developed and fitted to the reference data. Assuming that this correction is indeed transferable to the original molecules, one can use it not only to obtain an estimate for the total bias, but also to estimate the uncertainty of the final result (\textit{i.e.}, the result corrected for this bias). This approach has been applied recently by Pernot and others to estimate the uncertainties of different properties of solids, using a simple linear correction term\cite{Pernot2015}. Note also that the common scaling of vibrational frequencies (see, \textit{e.g.}, Refs.~\citenum{Scott1996, Irikura2005}) is nothing else but an attempt to correct the systematic error inherent in them. However, a key deficiency here is what ``molecular similarity'' actually means and how it can be assessed in a rigorous and objective manner.

This approach of removing systematic errors has the great advantage that it can potentially work with any kind of bias and, hence, it can be applied to any truly non-empiric quantum chemical method. However, it requires reference data which is essentially free of any errors---this requirement often severely hampers the application of this approach. Moreover, even when available, reference data can be inconsistent, which means that the standard deviation of repeated measurements is not consistent with the claimed uncertainty of the individual measurements. Recently, Pernot and Cailliez have published a review summarizing calibration procedure capable of dealing with inconsistent data\cite{Pernot2017}.

Most quantum chemical methods contain one or more empirical parameters which also introduce a certain error. During the past years, much progress has been done with respect to the estimation of such parametric uncertainties\cite{Simm2017}. Already in 2005, N\o{}rskov, Sethna, Jacobsen, and coworkers proposed a method to estimate errors of density functional calculations\cite{Mortensen2005}. Instead of only using the optimal parameters (as obtained from some fitting procedure) to evaluate a given exchange--correlation functional, an entire distribution of parameters is considered. This leads to an ensemble of density functionals, all of which give slightly different results for a particular system; this set of results can subsequently be used to obtain an estimate of the uncertainty inherent in this result. A similar approach has been presented by Aldegunde \textit{et al.}\cite{Aldegunde2016} and was extended to obtain system-specific density functionals including Bayesian uncertainty estimation\cite{Simm2016}.
Such approaches can over- or underestimate the prediction uncertainty. In many cases, the parameters of a certain model do implicitly depend on a control variable (like temperature). A model ensemble obtained for one value of the control variable does not necessarily provide correct uncertainty estimates for other values of the control variable\cite{Pernot2017a}.

Another option to consider parametric uncertainty is bootstrapping\cite{Efron1979}. In bootstrapping, starting from some data set, a superset of data sets is created by sampling with replacement. Every new data set has the same size as the original data set, but its exact composition will in general be different. Next, for a model whose parameters have been obtained for the original data set, the fitting procedure can be repeated for each of the new data sets, leading to a set of fitting parameters or, equivalently, to an ensemble of models. Again, this ensemble can be used to estimate the uncertainty of any particular result obtained with the model. 

Bootstrapping can be applied to quantify the error of physico-chemical models in general, as demonstrated at the examples of \textsuperscript{57}Fe M\"ossbauer isomer shift predictions \cite{Proppe2017} and of semiclassical dispersion corrections\cite{Weymuth2018}. 

An alternative to this Bayesian approach is to directly assign an upper and lower boundary on any parameter as is done in a methodology called ``data collaboration``\cite{Frenklach2002, Frenklach2004}. The core idea of this ansatz is to assemble information from different sources, both theoretical and experimental ones, in order to transfer experimental uncertainties, theoretical constraints, and prior bounds on model parameters directly into the model\cite{Russi2010}. This methodology is motivated by the fact that complex reaction networks, when assembled from elementary steps studied in isolation, usually do not yield a realistic model\cite{Frenklach1992}. The reason is that, in the full model, many parameters (and their uncertainties) are correlated with each other, and these correlations cannot be captured when only parts of the full model are studied. Therefore, instead of simply assembling a model from the best-fit parameters of its parts studied in isolation, one should directly take into account the full set of ``raw'' data in order to construct a model. Along these lines, Oreluk \textit{et al.}~applied a variant of data collaboration in the uncertainty quantification of the PM7 semiempirical model\cite{Oreluk2018}.

Another source of uncertainty in quantum chemical calculations is the molecular structure employed. While it is common that the sensitivity of a given result with respect to small structural variations is investigated, a framework to quantify the uncertainty due to structural sensitivity is largely missing. However, Jacob and coworkers presented interesting first steps into this direction for the calculation of theoretical spectra\cite{Oung2018, Bergmann2020}.

For models based on machine-learning techniques, uncertainty quantification is also of vital importance. Many ML models can provide a built-in measure of the expected uncertainty. Typically, the model reports a lower uncertainty for regions which are well represented in the training data set, and higher uncertainties for regions of chemical space which are outside the training set. This behavior can be exploited to construct a model which continuously improves itself as chemical space is explored (see above)\cite{Patra2017, Simm2018}. However, such uncertainty measures have to be interpreted with care. As an example, consider Gaussian processes. A Gaussian process is a collection of random variables which have a joint Gaussian distribution. That is, it is essentially a distribution over functions\cite{Rasmussen2004}. Every Gaussian process is defined by a mean function and a covariance function, which have to be given before training. The maximum uncertainty that a Gaussian process can report is defined by the covariance function. Therefore, outside the range of training, the Gaussian process essentially predicts the predefined mean function with an estimated uncertainty as given by the predefined covariance function. Obviously, this could be wrong. Therefore, additional ways to accurately quantify the uncertainty of a ML model need to be taken into consideration. In fact, Bayesian methods such as the bootstrap approach mentioned above can also applied in machine learning\cite{Peterson2017, Musil2019, Vishwakarma2021}. For example, Venturi \textit{et al.}~recently demonstrated how one can substitute the parameters in a neural net by a parameter distribution, thereby building a neural net for the prediction of potential energy surfaces with reliable uncertainty estimates\cite{Venturi2020}. However, such ensemble methods can lead to too low uncertainty estimates for molecules which are very dissimilar to any molecule present in the training data (so-called ``out-of-domain'' molecules)\cite{Liu2019}.

An alternative approach uses the distance between a molecule and the nearest training point (or a set of training points) as uncertainty estimate\cite{Liu2018, Janet2019}. To evaluate the distance, metrics such as the Tanimoto similarity\cite{Rogers1960} can be used. The core idea with such approaches is that the uncertainty of a prediction increases with decreasing similarity.

In addition to such approaches, standardized benchmarks are of great value because they allow for a straightforward comparison of different ML models. To this end, Brown and others have recently introduced a framework called ``GuacaMol''\cite{Brown2019}.

\section{Conclusions and Outlook}
\label{sec:conclusion}

The efficient search for new materials and molecules with custom-tailored properties is more important than ever, emphasizing the importance of
inverse quantum chemistry and rational design approaches.

The rational design of new compounds largely relies on three different strategies: The first is to build quantitative structure--property relationships in the hope to better understand how a given property depends on structural features. An inversion of such a relationship could then eventually allow to devise a molecule with certain desired properties. The second strategy is high-throughput virtual screening, \textit{i.e.}, to evaluate the properties of a large number of molecules in order to find target compounds. The third strategy is to take advantage of sophisticated sampling algorithms to be able to navigate chemical space efficiently while searching for a target compound. For all three strategies, machine learning has become to play a central role during the last decade, since machine learning models can often speed up the search for new compounds by several orders of magnitude.

Irrespective of all the technical advances which have been made over the past decades, chemical space is still so large that a full exploration is beyond any limits. Therefore, human chemical intuition and heuristics play a crucial role in inverse design. The selection of the chemical subspace to be searched as well as of the initial structures have a major impact, yet are far from trivial to be done algorithmically.

One aspect which has often been overlooked in inverse quantum chemistry is uncertainty quantification. Nevertheless, it is extremely important to have reliable error bars for any quantum chemical result in order to be able to properly judge its reliability. Even though the error of a given quantum chemical method is very difficult to reliably estimate due to the many different approximations which it incorporates, much progress has been made in the practical estimation of quantum chemical uncertainties during the past twenty years.

Today, the prospect of completely autonomous algorithms capable of searching huge parts of chemical space while continuously improving themselves is no longer an outlandish vision of a far-distant future. Although many challenges still exist, it can be expected that these will eventually be successfully addressed.

\section*{Acknowledgments}
\label{sec:acknowledgments}
This publication was created as part of NCCR Catalysis (grant number 180544), a National Centre of Competence in Research funded by the Swiss National Science Foundation.


\begin{mcitethebibliography}{143}
\providecommand*\natexlab[1]{#1}
\providecommand*\mciteSetBstSublistMode[1]{}
\providecommand*\mciteSetBstMaxWidthForm[2]{}
\providecommand*\mciteBstWouldAddEndPuncttrue
  {\def\EndOfBibitem{\unskip.}}
\providecommand*\mciteBstWouldAddEndPunctfalse
  {\let\EndOfBibitem\relax}
\providecommand*\mciteSetBstMidEndSepPunct[3]{}
\providecommand*\mciteSetBstSublistLabelBeginEnd[3]{}
\providecommand*\EndOfBibitem{}
\mciteSetBstSublistMode{f}
\mciteSetBstMaxWidthForm{subitem}{(\alph{mcitesubitemcount})}
\mciteSetBstSublistLabelBeginEnd
  {\mcitemaxwidthsubitemform\space}
  {\relax}
  {\relax}

\bibitem[Dykstra \latin{et~al.}(2005)Dykstra, Frenking, Kim, and
  Scuseria]{Dykstra2005}
Dykstra,~C.~E., Frenking,~G., Kim,~K.~S., Scuseria,~G.~E., Eds. \emph{{Theory
  and Applications of Computational Chemistry: The First Forty Years}};
  Elsevier: Amsterdam, 2005\relax
\mciteBstWouldAddEndPuncttrue
\mciteSetBstMidEndSepPunct{\mcitedefaultmidpunct}
{\mcitedefaultendpunct}{\mcitedefaultseppunct}\relax
\EndOfBibitem
\bibitem[Meuwly(2019)]{Meuwly2019}
Meuwly,~M. {Reactive Molecular Dynamics: From Small Molecules to Proteins}.
  \emph{WIREs Comput. Mol. Sci.} \textbf{2019}, \emph{9}, e1386\relax
\mciteBstWouldAddEndPuncttrue
\mciteSetBstMidEndSepPunct{\mcitedefaultmidpunct}
{\mcitedefaultendpunct}{\mcitedefaultseppunct}\relax
\EndOfBibitem
\bibitem[Dewyer \latin{et~al.}(2018)Dewyer, Arg{\"u}elles, and
  Zimmerman]{Dewyer2018}
Dewyer,~A.~L.; Arg{\"u}elles,~A.~J.; Zimmerman,~P.~M. {Methods for exploring
  reaction space in molecular systems}. \emph{Wiley Interdiscip. Rev. Comput.
  Mol. Sci.} \textbf{2018}, \emph{8}, e1354\relax
\mciteBstWouldAddEndPuncttrue
\mciteSetBstMidEndSepPunct{\mcitedefaultmidpunct}
{\mcitedefaultendpunct}{\mcitedefaultseppunct}\relax
\EndOfBibitem
\bibitem[V{\'a}zquez \latin{et~al.}(2018)V{\'a}zquez, Otero, and
  {Martinez-Nunez}]{Vazquez2018}
V{\'a}zquez,~S.~A.; Otero,~X.~L.; {Martinez-Nunez},~E. {A Trajectory-Based
  Method to Explore Reaction Mechanisms}. \emph{Molecules} \textbf{2018},
  \emph{23}, 3156\relax
\mciteBstWouldAddEndPuncttrue
\mciteSetBstMidEndSepPunct{\mcitedefaultmidpunct}
{\mcitedefaultendpunct}{\mcitedefaultseppunct}\relax
\EndOfBibitem
\bibitem[Simm \latin{et~al.}(2019)Simm, Vaucher, and Reiher]{Simm2019}
Simm,~G.~N.; Vaucher,~A.~C.; Reiher,~M. {Exploration of Reaction Pathways and
  Chemical Transformation Networks}. \emph{J. Phys. Chem. A} \textbf{2019},
  \emph{123}, 385--399\relax
\mciteBstWouldAddEndPuncttrue
\mciteSetBstMidEndSepPunct{\mcitedefaultmidpunct}
{\mcitedefaultendpunct}{\mcitedefaultseppunct}\relax
\EndOfBibitem
\bibitem[Unsleber and Reiher(2020)Unsleber, and Reiher]{Unsleber2020}
Unsleber,~J.~P.; Reiher,~M. {The Exploration of Chemical Reaction Networks}.
  \emph{Annu. Rev. Phys. Chem.} \textbf{2020}, \emph{71}, 121--142\relax
\mciteBstWouldAddEndPuncttrue
\mciteSetBstMidEndSepPunct{\mcitedefaultmidpunct}
{\mcitedefaultendpunct}{\mcitedefaultseppunct}\relax
\EndOfBibitem
\bibitem[Shannon \latin{et~al.}(2021)Shannon, {Mart{\'i}nez-N{\'u}{\~n}ez},
  Shalashilin, and Glowacki]{Shannon2021}
Shannon,~R.~J.; {Mart{\'i}nez-N{\'u}{\~n}ez},~E.; Shalashilin,~D.~V.;
  Glowacki,~D.~R. {ChemDyME: Kinetically Steered, Automated Mechanism
  Generation Through Combined Molecular Dynamics and Master Equation
  Calculations}. \emph{J. Chem. Theory Comput.} \textbf{2021}, \emph{17},
  4901--4912\relax
\mciteBstWouldAddEndPuncttrue
\mciteSetBstMidEndSepPunct{\mcitedefaultmidpunct}
{\mcitedefaultendpunct}{\mcitedefaultseppunct}\relax
\EndOfBibitem
\bibitem[Maeda and Harabuchi(2021)Maeda, and Harabuchi]{Maeda2021}
Maeda,~S.; Harabuchi,~Y. {Exploring paths of chemical transformations in
  molecular and periodic systems: An approach utilizing force}. \emph{Wiley
  Interdiscip. Rev. Comput. Mol. Sci.} \textbf{2021}, e1538\relax
\mciteBstWouldAddEndPuncttrue
\mciteSetBstMidEndSepPunct{\mcitedefaultmidpunct}
{\mcitedefaultendpunct}{\mcitedefaultseppunct}\relax
\EndOfBibitem
\bibitem[Steiner and Reiher(2022)Steiner, and Reiher]{Steiner2021}
Steiner,~M.; Reiher,~M. {Autonomous Reaction Network Exploration in Homogeneous
  and Heterogeneous Catalysis}. \emph{Top. Catal.} \textbf{2022}, \emph{65},
  6--39\relax
\mciteBstWouldAddEndPuncttrue
\mciteSetBstMidEndSepPunct{\mcitedefaultmidpunct}
{\mcitedefaultendpunct}{\mcitedefaultseppunct}\relax
\EndOfBibitem
\bibitem[Born and Oppenheimer(1927)Born, and Oppenheimer]{Born1927}
Born,~M.; Oppenheimer,~R. {Zur Quantentheorie der Molekeln}. \emph{Ann. Phys.}
  \textbf{1927}, \emph{389}, 457--484\relax
\mciteBstWouldAddEndPuncttrue
\mciteSetBstMidEndSepPunct{\mcitedefaultmidpunct}
{\mcitedefaultendpunct}{\mcitedefaultseppunct}\relax
\EndOfBibitem
\bibitem[Tully(2000)]{Tully2000}
Tully,~J.~C. {Perspective on ``Zur Quantentheorie der Molekeln''}. \emph{Theor.
  Chem. Acc.} \textbf{2000}, \emph{103}, 173--176\relax
\mciteBstWouldAddEndPuncttrue
\mciteSetBstMidEndSepPunct{\mcitedefaultmidpunct}
{\mcitedefaultendpunct}{\mcitedefaultseppunct}\relax
\EndOfBibitem
\bibitem[Norman \latin{et~al.}(2018)Norman, Ruud, and Saue]{Norman2018}
Norman,~P.; Ruud,~K.; Saue,~T. \emph{{Principles and Practices of Molecular
  Properties: Theory, Modeling and Simulations}}; John Wiley \& Sons Ltd:
  Oxford, 2018\relax
\mciteBstWouldAddEndPuncttrue
\mciteSetBstMidEndSepPunct{\mcitedefaultmidpunct}
{\mcitedefaultendpunct}{\mcitedefaultseppunct}\relax
\EndOfBibitem
\bibitem[Weymuth and Reiher(2014)Weymuth, and Reiher]{Weymuth2014a}
Weymuth,~T.; Reiher,~M. {Inverse Quantum Chemistry: Concepts and Strategies for
  Rational Compound Design}. \emph{Int. J. Quantum Chem.} \textbf{2014},
  \emph{114}, 823--837\relax
\mciteBstWouldAddEndPuncttrue
\mciteSetBstMidEndSepPunct{\mcitedefaultmidpunct}
{\mcitedefaultendpunct}{\mcitedefaultseppunct}\relax
\EndOfBibitem
\bibitem[Zunger(2018)]{Zunger2018}
Zunger,~A. {Inverse design in search of materials with target functionalities}.
  \emph{Nat. Rev. Chem.} \textbf{2018}, \emph{2}, 0121\relax
\mciteBstWouldAddEndPuncttrue
\mciteSetBstMidEndSepPunct{\mcitedefaultmidpunct}
{\mcitedefaultendpunct}{\mcitedefaultseppunct}\relax
\EndOfBibitem
\bibitem[Sabatier(2000)]{Sabatier2000}
Sabatier,~P.~C. {Past and future of inverse problems}. \emph{J. Math. Phys.}
  \textbf{2000}, \emph{41}, 4082--4124\relax
\mciteBstWouldAddEndPuncttrue
\mciteSetBstMidEndSepPunct{\mcitedefaultmidpunct}
{\mcitedefaultendpunct}{\mcitedefaultseppunct}\relax
\EndOfBibitem
\bibitem[Ruddigkeit \latin{et~al.}(2012)Ruddigkeit, {van Deursen}, Blum, and
  Reymond]{Ruddigkeit2012}
Ruddigkeit,~L.; {van Deursen},~R.; Blum,~L.~C.; Reymond,~J.-L. {Enumeration of
  166 Billion Organic Small Molecules in the Chemical Universe Database
  GDB-17}. \emph{J. Chem. Inf. Model.} \textbf{2012}, \emph{52},
  2864--2875\relax
\mciteBstWouldAddEndPuncttrue
\mciteSetBstMidEndSepPunct{\mcitedefaultmidpunct}
{\mcitedefaultendpunct}{\mcitedefaultseppunct}\relax
\EndOfBibitem
\bibitem[Reymond(2015)]{Reymond2015}
Reymond,~J.-L. {The Chemical Space Project}. \emph{Acc. Chem. Res.}
  \textbf{2015}, \emph{48}, 722--730\relax
\mciteBstWouldAddEndPuncttrue
\mciteSetBstMidEndSepPunct{\mcitedefaultmidpunct}
{\mcitedefaultendpunct}{\mcitedefaultseppunct}\relax
\EndOfBibitem
\bibitem[Bohacek \latin{et~al.}(1996)Bohacek, McMartin, and Guida]{Bohacek1996}
Bohacek,~R.~S.; McMartin,~C.; Guida,~W.~C. {The Art and Practice of
  Structure-Based Drug Design: A Molecular Modeling Perspective}. \emph{Med.
  Res. Rev.} \textbf{1996}, \emph{16}, 3--50\relax
\mciteBstWouldAddEndPuncttrue
\mciteSetBstMidEndSepPunct{\mcitedefaultmidpunct}
{\mcitedefaultendpunct}{\mcitedefaultseppunct}\relax
\EndOfBibitem
\bibitem[Dobson(2004)]{Dobson2004}
Dobson,~C.~M. {Chemical space and biology}. \emph{Nature} \textbf{2004},
  \emph{432}, 824--828\relax
\mciteBstWouldAddEndPuncttrue
\mciteSetBstMidEndSepPunct{\mcitedefaultmidpunct}
{\mcitedefaultendpunct}{\mcitedefaultseppunct}\relax
\EndOfBibitem
\bibitem[Freeze \latin{et~al.}(2019)Freeze, Kelly, and Batista]{Freeze2019}
Freeze,~J.~G.; Kelly,~H.~R.; Batista,~V.~S. {Search for Catalysts by Inverse
  Design: Artificial Intelligence, Mountain Climbers, and Alchemists}.
  \emph{Chem. Rev.} \textbf{2019}, \emph{119}, 6595--6612\relax
\mciteBstWouldAddEndPuncttrue
\mciteSetBstMidEndSepPunct{\mcitedefaultmidpunct}
{\mcitedefaultendpunct}{\mcitedefaultseppunct}\relax
\EndOfBibitem
\bibitem[Cros(1863)]{Cros1863}
Cros,~A. F.~A. {Action de l'alcohol amylique sur l'organisme}. Ph. D. thesis,
  University of Strasbourg, 1863\relax
\mciteBstWouldAddEndPuncttrue
\mciteSetBstMidEndSepPunct{\mcitedefaultmidpunct}
{\mcitedefaultendpunct}{\mcitedefaultseppunct}\relax
\EndOfBibitem
\bibitem[Hammett(1935)]{Hammett1935}
Hammett,~L.~P. {Some Relations between Reaction Rates and Equilibrium
  Constants.} \emph{Chem. Rev.} \textbf{1935}, \emph{17}, 125--136\relax
\mciteBstWouldAddEndPuncttrue
\mciteSetBstMidEndSepPunct{\mcitedefaultmidpunct}
{\mcitedefaultendpunct}{\mcitedefaultseppunct}\relax
\EndOfBibitem
\bibitem[Hammett(1937)]{Hammett1937}
Hammett,~L.~P. {The Effect of Structure upon the Reactions of Organic
  Compounds. Benzene Derivatives}. \emph{J. Am. Chem. Soc.} \textbf{1937},
  \emph{59}, 96--103\relax
\mciteBstWouldAddEndPuncttrue
\mciteSetBstMidEndSepPunct{\mcitedefaultmidpunct}
{\mcitedefaultendpunct}{\mcitedefaultseppunct}\relax
\EndOfBibitem
\bibitem[Hansch \latin{et~al.}(1962)Hansch, Maloney, Fujita, and
  Muir]{Hansch1962}
Hansch,~C.; Maloney,~P.~P.; Fujita,~T.; Muir,~R.~M. {Correlation of Biological
  Activity of Phenoxyacetic Acids with Hammett Substituent Constants and
  Partition Coefficients}. \emph{Nature} \textbf{1962}, \emph{194},
  178--180\relax
\mciteBstWouldAddEndPuncttrue
\mciteSetBstMidEndSepPunct{\mcitedefaultmidpunct}
{\mcitedefaultendpunct}{\mcitedefaultseppunct}\relax
\EndOfBibitem
\bibitem[Hansch \latin{et~al.}(1963)Hansch, Muir, Fujita, Maloney, Geiger, and
  Streich]{Hansch1963}
Hansch,~C.; Muir,~R.~M.; Fujita,~T.; Maloney,~P.~P.; Geiger,~F.; Streich,~M.
  {The Correlation of Biological Activity of Plant Growth Regulators and
  Chloromycetin Derivatives with Hammett Constants and Partition Coefficients}.
  \emph{J. Am. Chem. Soc.} \textbf{1963}, \emph{85}, 2817--2824\relax
\mciteBstWouldAddEndPuncttrue
\mciteSetBstMidEndSepPunct{\mcitedefaultmidpunct}
{\mcitedefaultendpunct}{\mcitedefaultseppunct}\relax
\EndOfBibitem
\bibitem[Hansch and Fujita(1964)Hansch, and Fujita]{Hansch1964}
Hansch,~C.; Fujita,~T. {{$\rho$}-{$\sigma$}-{$\pi$} Analysis. A Method for the
  Correlation of Biological Activity and Chemical Structure}. \emph{J. Am.
  Chem. Soc.} \textbf{1964}, \emph{86}, 1616--1626\relax
\mciteBstWouldAddEndPuncttrue
\mciteSetBstMidEndSepPunct{\mcitedefaultmidpunct}
{\mcitedefaultendpunct}{\mcitedefaultseppunct}\relax
\EndOfBibitem
\bibitem[Katritzky \latin{et~al.}(1995)Katritzky, Lobanov, and
  Karelson]{Katritzky1995}
Katritzky,~A.~R.; Lobanov,~V.~S.; Karelson,~M. {QSPR: The Correlation and
  Quantitative Prediction of Chemical and Physical Properties from Structure}.
  \emph{Chem. Soc. Rev.} \textbf{1995}, \emph{24}, 279--287\relax
\mciteBstWouldAddEndPuncttrue
\mciteSetBstMidEndSepPunct{\mcitedefaultmidpunct}
{\mcitedefaultendpunct}{\mcitedefaultseppunct}\relax
\EndOfBibitem
\bibitem[Karelson \latin{et~al.}(1996)Karelson, Lobanov, and
  Katritzky]{Karelson1996}
Karelson,~M.; Lobanov,~V.~S.; Katritzky,~A.~R. {Quantum-Chemical Descriptors in
  QSAR/QSPR Studies}. \emph{Chem. Rev.} \textbf{1996}, \emph{96},
  1027--1044\relax
\mciteBstWouldAddEndPuncttrue
\mciteSetBstMidEndSepPunct{\mcitedefaultmidpunct}
{\mcitedefaultendpunct}{\mcitedefaultseppunct}\relax
\EndOfBibitem
\bibitem[Katritzky \latin{et~al.}(2010)Katritzky, Kuanar, Slavov, Hall,
  Karelson, Kahn, and Dobchev]{Katritzky2010}
Katritzky,~A.~R.; Kuanar,~M.; Slavov,~S.; Hall,~C.~D.; Karelson,~M.; Kahn,~I.;
  Dobchev,~D.~A. {Quantitative Correlation of Physical and Chemical Properties
  with Chemical Structure: Utility for Prediction}. \emph{Chem. Rev.}
  \textbf{2010}, \emph{110}, 5714--5789\relax
\mciteBstWouldAddEndPuncttrue
\mciteSetBstMidEndSepPunct{\mcitedefaultmidpunct}
{\mcitedefaultendpunct}{\mcitedefaultseppunct}\relax
\EndOfBibitem
\bibitem[Le \latin{et~al.}(2012)Le, Epa, Burden, and Winkler]{Le2012}
Le,~T.; Epa,~V.~C.; Burden,~F.~R.; Winkler,~D.~A. {Quantitative
  Structure--Property Relationship Modeling of Diverse Materials Properties}.
  \emph{Chem. Rev.} \textbf{2012}, \emph{112}, 2889--2919\relax
\mciteBstWouldAddEndPuncttrue
\mciteSetBstMidEndSepPunct{\mcitedefaultmidpunct}
{\mcitedefaultendpunct}{\mcitedefaultseppunct}\relax
\EndOfBibitem
\bibitem[Berhanu \latin{et~al.}(2012)Berhanu, Pillai, Oliferenko, and
  Katritzky]{Berhanu2012}
Berhanu,~W.~M.; Pillai,~G.~G.; Oliferenko,~A.~A.; Katritzky,~A.~R.
  {Quantitative Structure--Activity/Property Relationships: The Ubiquitous
  Links between Cause and Effect}. \emph{ChemPlusChem} \textbf{2012},
  \emph{77}, 507--517\relax
\mciteBstWouldAddEndPuncttrue
\mciteSetBstMidEndSepPunct{\mcitedefaultmidpunct}
{\mcitedefaultendpunct}{\mcitedefaultseppunct}\relax
\EndOfBibitem
\bibitem[Pirhadi \latin{et~al.}(2015)Pirhadi, Shiri, and Ghasemi]{Pirhadi2015}
Pirhadi,~S.; Shiri,~F.; Ghasemi,~J.~B. {Multivariate statistical analysis
  methods in QSAR}. \emph{RSC Adv.} \textbf{2015}, \emph{5},
  104635--104665\relax
\mciteBstWouldAddEndPuncttrue
\mciteSetBstMidEndSepPunct{\mcitedefaultmidpunct}
{\mcitedefaultendpunct}{\mcitedefaultseppunct}\relax
\EndOfBibitem
\bibitem[Polishchuk(2017)]{Polishchuk2017}
Polishchuk,~P. {Interpretation of Quantitative Structure--Activity Relationship
  Models: Past, Present, and Future}. \emph{J. Chem. Inf. Model.}
  \textbf{2017}, \emph{57}, 2618--2639\relax
\mciteBstWouldAddEndPuncttrue
\mciteSetBstMidEndSepPunct{\mcitedefaultmidpunct}
{\mcitedefaultendpunct}{\mcitedefaultseppunct}\relax
\EndOfBibitem
\bibitem[Muratov \latin{et~al.}(2020)Muratov, Bajorath, Sheridan, Tetko,
  Filimonov, Poroikov, Oprea, Baskin, Varnek, Roitberg, Isayev, Curtalolo,
  Fourches, Cohen, {Aspuru-Guzik}, Winkler, Agrafiotis, Cherkasov, and
  Tropsha]{Muratov2020}
Muratov,~E.~N. \latin{et~al.}  {QSAR without borders}. \emph{Chem. Soc. Rev.}
  \textbf{2020}, \emph{49}, 3525--3564\relax
\mciteBstWouldAddEndPuncttrue
\mciteSetBstMidEndSepPunct{\mcitedefaultmidpunct}
{\mcitedefaultendpunct}{\mcitedefaultseppunct}\relax
\EndOfBibitem
\bibitem[Rogers and Tanimoto(1960)Rogers, and Tanimoto]{Rogers1960}
Rogers,~D.~J.; Tanimoto,~T.~T. {A Computer Program for Classifying Plants}.
  \emph{Science} \textbf{1960}, \emph{132}, 1115--1118\relax
\mciteBstWouldAddEndPuncttrue
\mciteSetBstMidEndSepPunct{\mcitedefaultmidpunct}
{\mcitedefaultendpunct}{\mcitedefaultseppunct}\relax
\EndOfBibitem
\bibitem[Nikolova and Jaworska(2003)Nikolova, and Jaworska]{Nikolova2003}
Nikolova,~N.; Jaworska,~J. {Approaches to Measure Chemical Similarity -- a
  Review}. \emph{QSAR Comb. Sci.} \textbf{2003}, \emph{22}, 1006--1026\relax
\mciteBstWouldAddEndPuncttrue
\mciteSetBstMidEndSepPunct{\mcitedefaultmidpunct}
{\mcitedefaultendpunct}{\mcitedefaultseppunct}\relax
\EndOfBibitem
\bibitem[Fujita and Winkler(2016)Fujita, and Winkler]{Fujita2016}
Fujita,~T.; Winkler,~D.~A. {Understanding the Roles of the ``Two QSARs''}.
  \emph{J. Chem. Inf. Model.} \textbf{2016}, \emph{56}, 269--274\relax
\mciteBstWouldAddEndPuncttrue
\mciteSetBstMidEndSepPunct{\mcitedefaultmidpunct}
{\mcitedefaultendpunct}{\mcitedefaultseppunct}\relax
\EndOfBibitem
\bibitem[Rothenberg(2008)]{Rothenberg2008}
Rothenberg,~G. \emph{{Catalysis: Concepts and Green Applications}}; Wiley-VCH:
  Weinheim, 2008\relax
\mciteBstWouldAddEndPuncttrue
\mciteSetBstMidEndSepPunct{\mcitedefaultmidpunct}
{\mcitedefaultendpunct}{\mcitedefaultseppunct}\relax
\EndOfBibitem
\bibitem[Groom \latin{et~al.}(2016)Groom, Bruno, Lightfoot, and
  Ward]{Groom2016}
Groom,~C.~R.; Bruno,~I.~J.; Lightfoot,~M.~P.; Ward,~S.~C. {The Cambridge
  Structural Database}. \emph{Acta Cryst. B} \textbf{2016}, \emph{72},
  171--179\relax
\mciteBstWouldAddEndPuncttrue
\mciteSetBstMidEndSepPunct{\mcitedefaultmidpunct}
{\mcitedefaultendpunct}{\mcitedefaultseppunct}\relax
\EndOfBibitem
\bibitem[Zhang \latin{et~al.}(2019)Zhang, Chen, Su, and Li]{Zhang2019}
Zhang,~L.; Chen,~Z.; Su,~J.; Li,~J. {Data mining new energy materials from
  structure databases}. \emph{Renewable Sustainable Energy Rev.} \textbf{2019},
  \emph{107}, 554--567\relax
\mciteBstWouldAddEndPuncttrue
\mciteSetBstMidEndSepPunct{\mcitedefaultmidpunct}
{\mcitedefaultendpunct}{\mcitedefaultseppunct}\relax
\EndOfBibitem
\bibitem[Miyao \latin{et~al.}(2016)Miyao, Kaneko, and Funatsu]{Miyao2016}
Miyao,~T.; Kaneko,~H.; Funatsu,~K. {Inverse QSPR/QSAR Analysis for Chemical
  Structure Generation (from y to x)}. \emph{J. Chem. Inf. Model.}
  \textbf{2016}, \emph{56}, 286--299\relax
\mciteBstWouldAddEndPuncttrue
\mciteSetBstMidEndSepPunct{\mcitedefaultmidpunct}
{\mcitedefaultendpunct}{\mcitedefaultseppunct}\relax
\EndOfBibitem
\bibitem[Zheng \latin{et~al.}(1998)Zheng, Cho, and Tropsha]{Zheng1998}
Zheng,~W.; Cho,~S.~J.; Tropsha,~A. {Rational Combinatorial Library Design. 1.
  Focus-2D:\, A New Approach to the Design of Targeted Combinatorial Chemical
  Libraries}. \emph{J. Chem. Inf. Comput. Sci.} \textbf{1998}, \emph{38},
  251--258\relax
\mciteBstWouldAddEndPuncttrue
\mciteSetBstMidEndSepPunct{\mcitedefaultmidpunct}
{\mcitedefaultendpunct}{\mcitedefaultseppunct}\relax
\EndOfBibitem
\bibitem[Cho \latin{et~al.}(1998)Cho, Zheng, and Tropsha]{Cho1998}
Cho,~S.~J.; Zheng,~W.; Tropsha,~A. {Rational Combinatorial Library Design. 2.
  Rational Design of Targeted Combinatorial Peptide Libraries Using Chemical
  Similarity Probe and the Inverse QSAR Approaches}. \emph{J. Chem. Inf.
  Comput. Sci.} \textbf{1998}, \emph{38}, 259--268\relax
\mciteBstWouldAddEndPuncttrue
\mciteSetBstMidEndSepPunct{\mcitedefaultmidpunct}
{\mcitedefaultendpunct}{\mcitedefaultseppunct}\relax
\EndOfBibitem
\bibitem[Gantzer \latin{et~al.}(2020)Gantzer, Creton, and
  {Nieto-Draghi}]{Gantzer2020}
Gantzer,~P.; Creton,~B.; {Nieto-Draghi},~C. {Inverse-QSPR for {\emph{de novo}}
  Design: A Review}. \emph{Mol. Inf.} \textbf{2020}, \emph{39}, 1900087\relax
\mciteBstWouldAddEndPuncttrue
\mciteSetBstMidEndSepPunct{\mcitedefaultmidpunct}
{\mcitedefaultendpunct}{\mcitedefaultseppunct}\relax
\EndOfBibitem
\bibitem[Shoichet(2004)]{Shoichet2004}
Shoichet,~B.~K. {Virtual screening of chemical libraries}. \emph{Nature}
  \textbf{2004}, \emph{432}, 862--865\relax
\mciteBstWouldAddEndPuncttrue
\mciteSetBstMidEndSepPunct{\mcitedefaultmidpunct}
{\mcitedefaultendpunct}{\mcitedefaultseppunct}\relax
\EndOfBibitem
\bibitem[{Pyzer-Knapp} \latin{et~al.}(2015){Pyzer-Knapp}, Suh,
  {G{\'o}mez-Bombarelli}, {Aguilera-Iparraguirre}, and
  {Aspuru-Guzik}]{Pyzer-Knapp2015}
{Pyzer-Knapp},~E.~O.; Suh,~C.; {G{\'o}mez-Bombarelli},~R.;
  {Aguilera-Iparraguirre},~J.; {Aspuru-Guzik},~A. {What Is High-Throughput
  Virtual Screening? A Perspective from Organic Materials Discovery}.
  \emph{Annu. Rev. Mater. Res.} \textbf{2015}, \emph{45}, 195--216\relax
\mciteBstWouldAddEndPuncttrue
\mciteSetBstMidEndSepPunct{\mcitedefaultmidpunct}
{\mcitedefaultendpunct}{\mcitedefaultseppunct}\relax
\EndOfBibitem
\bibitem[{Attene-Ramos} \latin{et~al.}(2014){Attene-Ramos}, Austin, and
  Xia]{Attene-Ramos2014}
{Attene-Ramos},~M.~S.; Austin,~C.~P.; Xia,~M. In \emph{{Encyclopedia of
  Toxicology}}, 3rd ed.; Wexler,~P., Ed.; Academic Press: Oxford, 2014; Vol.~2;
  pp 916--917\relax
\mciteBstWouldAddEndPuncttrue
\mciteSetBstMidEndSepPunct{\mcitedefaultmidpunct}
{\mcitedefaultendpunct}{\mcitedefaultseppunct}\relax
\EndOfBibitem
\bibitem[Bevan \latin{et~al.}(1995)Bevan, Ryder, and Shaw]{Bevan1995}
Bevan,~P.; Ryder,~H.; Shaw,~I. {Identifying small-molecule lead compounds: The
  screening approach to drug discovery}. \emph{Trends Biotechnol.}
  \textbf{1995}, \emph{13}, 115--121\relax
\mciteBstWouldAddEndPuncttrue
\mciteSetBstMidEndSepPunct{\mcitedefaultmidpunct}
{\mcitedefaultendpunct}{\mcitedefaultseppunct}\relax
\EndOfBibitem
\bibitem[Hachmann \latin{et~al.}(2011)Hachmann, {Olivares-Amaya},
  {Atahan-Evrenk}, {Amador-Bedolla}, {S{\'a}nchez-Carrera}, {Gold-Parker},
  Vogt, Brockway, and {Aspuru-Guzik}]{Hachmann2011}
Hachmann,~J.; {Olivares-Amaya},~R.; {Atahan-Evrenk},~S.; {Amador-Bedolla},~C.;
  {S{\'a}nchez-Carrera},~R.~S.; {Gold-Parker},~A.; Vogt,~L.; Brockway,~A.~M.;
  {Aspuru-Guzik},~A. {The Harvard Clean Energy Project: Large-Scale
  Computational Screening and Design of Organic Photovoltaics on the World
  Community Grid}. \emph{J. Phys. Chem. Lett.} \textbf{2011}, \emph{2},
  2241--2251\relax
\mciteBstWouldAddEndPuncttrue
\mciteSetBstMidEndSepPunct{\mcitedefaultmidpunct}
{\mcitedefaultendpunct}{\mcitedefaultseppunct}\relax
\EndOfBibitem
\bibitem[Hachmann \latin{et~al.}(2014)Hachmann, {Olivares-Amaya}, Jinich,
  Appleton, {Blood-Forsythe}, Seress, {Rom{\'a}n-Salgado}, Trepte,
  {Atahan-Evrenk}, Er, Shrestha, Mondal, Sokolov, Bao, and
  {Aspuru-Guzik}]{Hachmann2014}
Hachmann,~J.; {Olivares-Amaya},~R.; Jinich,~A.; Appleton,~A.~L.;
  {Blood-Forsythe},~M.~A.; Seress,~L.~R.; {Rom{\'a}n-Salgado},~C.; Trepte,~K.;
  {Atahan-Evrenk},~S.; Er,~S.; Shrestha,~S.; Mondal,~R.; Sokolov,~A.; Bao,~Z.;
  {Aspuru-Guzik},~A. {Lead candidates for high-performance organic
  photovoltaics from high-throughput quantum chemistry -- the Harvard Clean
  Energy Project}. \emph{Energy Environ. Sci.} \textbf{2014}, \emph{7},
  698--704\relax
\mciteBstWouldAddEndPuncttrue
\mciteSetBstMidEndSepPunct{\mcitedefaultmidpunct}
{\mcitedefaultendpunct}{\mcitedefaultseppunct}\relax
\EndOfBibitem
\bibitem[Clery(2005)]{Clery2005}
Clery,~D. {IBM Offers Free Number Crunching for Humanitarian Research
  Projects}. \emph{Science} \textbf{2005}, \emph{308}, 773--773\relax
\mciteBstWouldAddEndPuncttrue
\mciteSetBstMidEndSepPunct{\mcitedefaultmidpunct}
{\mcitedefaultendpunct}{\mcitedefaultseppunct}\relax
\EndOfBibitem
\bibitem[Pollice \latin{et~al.}(2021)Pollice, {dos Passos Gomes}, Aldeghi,
  Hickman, Krenn, Lavigne, {Lindner-D'Addario}, Nigam, Ser, Yao, and
  {Aspuru-Guzik}]{Pollice2021}
Pollice,~R.; {dos Passos Gomes},~G.; Aldeghi,~M.; Hickman,~R.~J.; Krenn,~M.;
  Lavigne,~C.; {Lindner-D'Addario},~M.; Nigam,~A.; Ser,~C.~T.; Yao,~Z.;
  {Aspuru-Guzik},~A. {Data-Driven Strategies for Accelerated Materials Design}.
  \emph{Acc. Chem. Res.} \textbf{2021}, \emph{54}, 849--860\relax
\mciteBstWouldAddEndPuncttrue
\mciteSetBstMidEndSepPunct{\mcitedefaultmidpunct}
{\mcitedefaultendpunct}{\mcitedefaultseppunct}\relax
\EndOfBibitem
\bibitem[Meredig and Wolverton(2014)Meredig, and Wolverton]{Meredig2014}
Meredig,~B.; Wolverton,~C. {Dissolving the Periodic Table in Cubic Zirconia:
  Data Mining to Discover Chemical Trends}. \emph{Chem. Mater.} \textbf{2014},
  \emph{26}, 1985--1991\relax
\mciteBstWouldAddEndPuncttrue
\mciteSetBstMidEndSepPunct{\mcitedefaultmidpunct}
{\mcitedefaultendpunct}{\mcitedefaultseppunct}\relax
\EndOfBibitem
\bibitem[Kirkpatrick \latin{et~al.}(1983)Kirkpatrick, Gelatt, and
  Vecchi]{Kirkpatrick1983}
Kirkpatrick,~S.; Gelatt,~C.~D.,~Jr.; Vecchi,~M.~P. {Optimization by Simulated
  Annealing}. \emph{Science} \textbf{1983}, \emph{220}, 671--680\relax
\mciteBstWouldAddEndPuncttrue
\mciteSetBstMidEndSepPunct{\mcitedefaultmidpunct}
{\mcitedefaultendpunct}{\mcitedefaultseppunct}\relax
\EndOfBibitem
\bibitem[Kirkpatrick(1984)]{Kirkpatrick1984}
Kirkpatrick,~S. {Optimization by Simulated Annealing: Quantitative Studies}.
  \emph{J. Stat. Phys.} \textbf{1984}, \emph{34}, 975--986\relax
\mciteBstWouldAddEndPuncttrue
\mciteSetBstMidEndSepPunct{\mcitedefaultmidpunct}
{\mcitedefaultendpunct}{\mcitedefaultseppunct}\relax
\EndOfBibitem
\bibitem[Sch{\"o}n and Jansen(1996)Sch{\"o}n, and Jansen]{Schon1996}
Sch{\"o}n,~J.~C.; Jansen,~M. {First Step Towards Planning of Syntheses in
  Solid-State Chemistry: Determination of Promising Structure Candidates by
  Global Optimization}. \emph{Angew. Chem. Int. Ed.} \textbf{1996}, \emph{35},
  1286--1304\relax
\mciteBstWouldAddEndPuncttrue
\mciteSetBstMidEndSepPunct{\mcitedefaultmidpunct}
{\mcitedefaultendpunct}{\mcitedefaultseppunct}\relax
\EndOfBibitem
\bibitem[Jansen(2002)]{Jansen2002}
Jansen,~M. {A Concept for Synthesis Planning in Solid-State Chemistry}.
  \emph{Angew. Chem. Int. Ed.} \textbf{2002}, \emph{41}, 3746--3766\relax
\mciteBstWouldAddEndPuncttrue
\mciteSetBstMidEndSepPunct{\mcitedefaultmidpunct}
{\mcitedefaultendpunct}{\mcitedefaultseppunct}\relax
\EndOfBibitem
\bibitem[Franceschetti and Zunger(1999)Franceschetti, and
  Zunger]{Franceschetti1999}
Franceschetti,~A.; Zunger,~A. {The inverse band-structure problem of finding an
  atomic configuration with given electronic properties}. \emph{Nature}
  \textbf{1999}, \emph{402}, 60--63\relax
\mciteBstWouldAddEndPuncttrue
\mciteSetBstMidEndSepPunct{\mcitedefaultmidpunct}
{\mcitedefaultendpunct}{\mcitedefaultseppunct}\relax
\EndOfBibitem
\bibitem[Dudiy and Zunger(2006)Dudiy, and Zunger]{Dudiy2006}
Dudiy,~S.~V.; Zunger,~A. {Searching for Alloy Configurations with Target
  Physical Properties: Impurity Design via a Genetic Algorithm Inverse Band
  Structure Approach}. \emph{Phys. Rev. Lett.} \textbf{2006}, \emph{97},
  046401\relax
\mciteBstWouldAddEndPuncttrue
\mciteSetBstMidEndSepPunct{\mcitedefaultmidpunct}
{\mcitedefaultendpunct}{\mcitedefaultseppunct}\relax
\EndOfBibitem
\bibitem[{d'Avezac} \latin{et~al.}(2012){d'Avezac}, Luo, Chanier, and
  Zunger]{Avezac2012}
{d'Avezac},~M.; Luo,~J.-W.; Chanier,~T.; Zunger,~A. {Genetic-Algorithm
  Discovery of a Direct-Gap and Optically Allowed Superstructure from
  Indirect-Gap Si and Ge Semiconductors}. \emph{Phys. Rev. Lett.}
  \textbf{2012}, \emph{108}, 027401\relax
\mciteBstWouldAddEndPuncttrue
\mciteSetBstMidEndSepPunct{\mcitedefaultmidpunct}
{\mcitedefaultendpunct}{\mcitedefaultseppunct}\relax
\EndOfBibitem
\bibitem[Venkatasubramanian \latin{et~al.}(1994)Venkatasubramanian, Chan, and
  Caruthers]{Venkatasubramanian1994}
Venkatasubramanian,~V.; Chan,~K.; Caruthers,~J.~M. {Computer-aided molecular
  design using genetic algorithms}. \emph{Comput. Chem. Eng.} \textbf{1994},
  \emph{18}, 833--844\relax
\mciteBstWouldAddEndPuncttrue
\mciteSetBstMidEndSepPunct{\mcitedefaultmidpunct}
{\mcitedefaultendpunct}{\mcitedefaultseppunct}\relax
\EndOfBibitem
\bibitem[Le and Winkler(2016)Le, and Winkler]{Le2016}
Le,~T.~C.; Winkler,~D.~A. {Discovery and Optimization of Materials Using
  Evolutionary Approaches}. \emph{Chem. Rev.} \textbf{2016}, \emph{116},
  6107--6132\relax
\mciteBstWouldAddEndPuncttrue
\mciteSetBstMidEndSepPunct{\mcitedefaultmidpunct}
{\mcitedefaultendpunct}{\mcitedefaultseppunct}\relax
\EndOfBibitem
\bibitem[Damewood \latin{et~al.}(2010)Damewood, Lerman, and
  Masek]{Damewood2010}
Damewood,~J.~R.,~Jr.; Lerman,~C.~L.; Masek,~B.~B. {NovoFLAP: A Ligand-Based De
  Novo Design Approach for the Generation of Medicinally Relevant Ideas}.
  \emph{J. Chem. Inf. Model.} \textbf{2010}, \emph{50}, 1296--1303\relax
\mciteBstWouldAddEndPuncttrue
\mciteSetBstMidEndSepPunct{\mcitedefaultmidpunct}
{\mcitedefaultendpunct}{\mcitedefaultseppunct}\relax
\EndOfBibitem
\bibitem[Virshup \latin{et~al.}(2013)Virshup, {Contreras-Garc{\'i}a}, Wipf,
  Yang, and Beratan]{Virshup2013}
Virshup,~A.~M.; {Contreras-Garc{\'i}a},~J.; Wipf,~P.; Yang,~W.; Beratan,~D.~N.
  {Stochastic Voyages into Uncharted Chemical Space Produce a Representative
  Library of All Possible Drug-Like Compounds}. \emph{J. Am. Chem. Soc.}
  \textbf{2013}, \emph{135}, 7296--7303\relax
\mciteBstWouldAddEndPuncttrue
\mciteSetBstMidEndSepPunct{\mcitedefaultmidpunct}
{\mcitedefaultendpunct}{\mcitedefaultseppunct}\relax
\EndOfBibitem
\bibitem[Rupakheti \latin{et~al.}(2015)Rupakheti, Virshup, Yang, and
  Beratan]{Rupakheti2015}
Rupakheti,~C.; Virshup,~A.; Yang,~W.; Beratan,~D.~N. {Strategy To Discover
  Diverse Optimal Molecules in the Small Molecule Universe}. \emph{J. Chem.
  Inf. Model.} \textbf{2015}, \emph{55}, 529--537\relax
\mciteBstWouldAddEndPuncttrue
\mciteSetBstMidEndSepPunct{\mcitedefaultmidpunct}
{\mcitedefaultendpunct}{\mcitedefaultseppunct}\relax
\EndOfBibitem
\bibitem[Springborg \latin{et~al.}(2017)Springborg, Kohaut, Dong, and
  Huwig]{Springborg2017}
Springborg,~M.; Kohaut,~S.; Dong,~Y.; Huwig,~K. {Mixed Si-Ge clusters,
  solar-energy harvesting, and inverse-design methods}. \emph{Comp. Theor.
  Chem.} \textbf{2017}, \emph{1107}, 14--22\relax
\mciteBstWouldAddEndPuncttrue
\mciteSetBstMidEndSepPunct{\mcitedefaultmidpunct}
{\mcitedefaultendpunct}{\mcitedefaultseppunct}\relax
\EndOfBibitem
\bibitem[Huwig \latin{et~al.}(2017)Huwig, Fan, and Springborg]{Huwig2017}
Huwig,~K.; Fan,~C.; Springborg,~M. {From properties to materials: An efficient
  and simple approach}. \emph{J. Chem. Phys.} \textbf{2017}, \emph{147},
  234105\relax
\mciteBstWouldAddEndPuncttrue
\mciteSetBstMidEndSepPunct{\mcitedefaultmidpunct}
{\mcitedefaultendpunct}{\mcitedefaultseppunct}\relax
\EndOfBibitem
\bibitem[Nicolaou \latin{et~al.}(2009)Nicolaou, Apostolakis, and
  Pattichis]{Nicolaou2009}
Nicolaou,~C.~A.; Apostolakis,~J.; Pattichis,~C.~S. {De Novo Drug Design Using
  Multiobjective Evolutionary Graphs}. \emph{J. Chem. Inf. Model.}
  \textbf{2009}, \emph{49}, 295--307\relax
\mciteBstWouldAddEndPuncttrue
\mciteSetBstMidEndSepPunct{\mcitedefaultmidpunct}
{\mcitedefaultendpunct}{\mcitedefaultseppunct}\relax
\EndOfBibitem
\bibitem[Foscato \latin{et~al.}(2019)Foscato, Venkatraman, and
  Jensen]{Foscato2019}
Foscato,~M.; Venkatraman,~V.; Jensen,~V.~R. {DENOPTIM: Software for
  Computational {\emph{de Novo}} Design of Organic and Inorganic Molecules}.
  \emph{J. Chem. Inf. Model.} \textbf{2019}, \emph{59}, 4077--4082\relax
\mciteBstWouldAddEndPuncttrue
\mciteSetBstMidEndSepPunct{\mcitedefaultmidpunct}
{\mcitedefaultendpunct}{\mcitedefaultseppunct}\relax
\EndOfBibitem
\bibitem[Lameijer \latin{et~al.}(2006)Lameijer, Kok, B{\"a}ck, and
  IJzerman]{Lameijer2006}
Lameijer,~E.-W.; Kok,~J.~N.; B{\"a}ck,~T.; IJzerman,~A.~P. {The Molecule
  Evoluator. An Interactive Evolutionary Algorithm for the Design of Drug-Like
  Molecules}. \emph{J. Chem. Inf. Model.} \textbf{2006}, \emph{46},
  545--552\relax
\mciteBstWouldAddEndPuncttrue
\mciteSetBstMidEndSepPunct{\mcitedefaultmidpunct}
{\mcitedefaultendpunct}{\mcitedefaultseppunct}\relax
\EndOfBibitem
\bibitem[Kawai \latin{et~al.}(2014)Kawai, Nagata, and Takahashi]{Kawai2014}
Kawai,~K.; Nagata,~N.; Takahashi,~Y. {De Novo Design of Drug-Like Molecules by
  a Fragment-Based Molecular Evolutionary Approach}. \emph{J. Chem. Inf.
  Model.} \textbf{2014}, \emph{54}, 49--56\relax
\mciteBstWouldAddEndPuncttrue
\mciteSetBstMidEndSepPunct{\mcitedefaultmidpunct}
{\mcitedefaultendpunct}{\mcitedefaultseppunct}\relax
\EndOfBibitem
\bibitem[Fechner and Schneider(2006)Fechner, and Schneider]{Fechner2006}
Fechner,~U.; Schneider,~G. {Flux (1):\, A Virtual Synthesis Scheme for
  Fragment-Based de Novo Design}. \emph{J. Chem. Inf. Model.} \textbf{2006},
  \emph{46}, 699--707\relax
\mciteBstWouldAddEndPuncttrue
\mciteSetBstMidEndSepPunct{\mcitedefaultmidpunct}
{\mcitedefaultendpunct}{\mcitedefaultseppunct}\relax
\EndOfBibitem
\bibitem[Fechner and Schneider(2007)Fechner, and Schneider]{Fechner2007}
Fechner,~U.; Schneider,~G. {Flux (2):\, Comparison of Molecular Mutation and
  Crossover Operators for Ligand-Based de Novo Design}. \emph{J. Chem. Inf.
  Model.} \textbf{2007}, \emph{47}, 656--667\relax
\mciteBstWouldAddEndPuncttrue
\mciteSetBstMidEndSepPunct{\mcitedefaultmidpunct}
{\mcitedefaultendpunct}{\mcitedefaultseppunct}\relax
\EndOfBibitem
\bibitem[Pearl and Korf(1987)Pearl, and Korf]{Pearl1987}
Pearl,~J.; Korf,~R.~E. {Search Techniques}. \emph{Annu. Rev. Comput. Sci.}
  \textbf{1987}, \emph{2}, 451--467\relax
\mciteBstWouldAddEndPuncttrue
\mciteSetBstMidEndSepPunct{\mcitedefaultmidpunct}
{\mcitedefaultendpunct}{\mcitedefaultseppunct}\relax
\EndOfBibitem
\bibitem[De~Vleeschouwer \latin{et~al.}(2012)De~Vleeschouwer, Yang, Beratan,
  Geerlings, and De~Proft]{DeVleeschouwer2012}
De~Vleeschouwer,~F.; Yang,~W.; Beratan,~D.~N.; Geerlings,~P.; De~Proft,~F.
  {Inverse design of molecules with optimal reactivity properties: acidity of
  2-naphthol derivatives}. \emph{Phys. Chem. Chem. Phys.} \textbf{2012},
  \emph{14}, 16002--16013\relax
\mciteBstWouldAddEndPuncttrue
\mciteSetBstMidEndSepPunct{\mcitedefaultmidpunct}
{\mcitedefaultendpunct}{\mcitedefaultseppunct}\relax
\EndOfBibitem
\bibitem[De~Vleeschouwer \latin{et~al.}(2016)De~Vleeschouwer, Geerlings, and
  De~Proft]{DeVleeschouwer2016}
De~Vleeschouwer,~F.; Geerlings,~P.; De~Proft,~F. {Molecular Property
  Optimizations with Boundary Conditions through the Best First Search Scheme}.
  \emph{ChemPhysChem} \textbf{2016}, \emph{17}, 1414--1424\relax
\mciteBstWouldAddEndPuncttrue
\mciteSetBstMidEndSepPunct{\mcitedefaultmidpunct}
{\mcitedefaultendpunct}{\mcitedefaultseppunct}\relax
\EndOfBibitem
\bibitem[Weymuth and Reiher(2014)Weymuth, and Reiher]{Weymuth2014b}
Weymuth,~T.; Reiher,~M. {Gradient-Driven Molecule Construction: An Inverse
  Approach Applied to the Design of Small-Molecule Fixating Catalysts}.
  \emph{Int. J. Quantum Chem.} \textbf{2014}, \emph{114}, 838--850\relax
\mciteBstWouldAddEndPuncttrue
\mciteSetBstMidEndSepPunct{\mcitedefaultmidpunct}
{\mcitedefaultendpunct}{\mcitedefaultseppunct}\relax
\EndOfBibitem
\bibitem[Weymuth and Reiher(2013)Weymuth, and Reiher]{Weymuth2013}
Weymuth,~T.; Reiher,~M. {Toward an Inverse Approach for the Design of
  Small-Molecule Fixating Catalysts}. \emph{MRS Proc.} \textbf{2013},
  \emph{1524}, 601\relax
\mciteBstWouldAddEndPuncttrue
\mciteSetBstMidEndSepPunct{\mcitedefaultmidpunct}
{\mcitedefaultendpunct}{\mcitedefaultseppunct}\relax
\EndOfBibitem
\bibitem[Krausbeck \latin{et~al.}(2017)Krausbeck, Sobez, and
  Reiher]{Krausbeck2017}
Krausbeck,~F.; Sobez,~J.-G.; Reiher,~M. {Stabilization of Activated Fragments
  by Shell-Wise Construction of an Embedding Environment}. \emph{J. Comput.
  Chem.} \textbf{2017}, \emph{38}, 1023--1038\relax
\mciteBstWouldAddEndPuncttrue
\mciteSetBstMidEndSepPunct{\mcitedefaultmidpunct}
{\mcitedefaultendpunct}{\mcitedefaultseppunct}\relax
\EndOfBibitem
\bibitem[Ikebata \latin{et~al.}(2017)Ikebata, Hongo, Isomura, Maezono, and
  Yoshida]{Ikebata2017}
Ikebata,~H.; Hongo,~K.; Isomura,~T.; Maezono,~R.; Yoshida,~R. {Bayesian
  molecular design with a chemical language model}. \emph{J. Comput. Aided Mol.
  Des.} \textbf{2017}, \emph{31}, 379--391\relax
\mciteBstWouldAddEndPuncttrue
\mciteSetBstMidEndSepPunct{\mcitedefaultmidpunct}
{\mcitedefaultendpunct}{\mcitedefaultseppunct}\relax
\EndOfBibitem
\bibitem[Dittner and Hartke(2018)Dittner, and Hartke]{Dittner2018}
Dittner,~M.; Hartke,~B. {Globally Optimal Catalytic Fields -- Inverse Design of
  Abstract Embeddings for Maximum Reaction Rate Acceleration}. \emph{J. Chem.
  Theory Comput.} \textbf{2018}, \emph{14}, 3547--3564\relax
\mciteBstWouldAddEndPuncttrue
\mciteSetBstMidEndSepPunct{\mcitedefaultmidpunct}
{\mcitedefaultendpunct}{\mcitedefaultseppunct}\relax
\EndOfBibitem
\bibitem[Dittner and Hartke(2020)Dittner, and Hartke]{Dittner2020}
Dittner,~M.; Hartke,~B. {Globally optimal catalytic fields for a Diels--Alder
  reaction}. \emph{J. Chem. Phys.} \textbf{2020}, \emph{152}, 114106\relax
\mciteBstWouldAddEndPuncttrue
\mciteSetBstMidEndSepPunct{\mcitedefaultmidpunct}
{\mcitedefaultendpunct}{\mcitedefaultseppunct}\relax
\EndOfBibitem
\bibitem[Behrens and Hartke(2021)Behrens, and Hartke]{Behrens2021}
Behrens,~D.~M.; Hartke,~B. {Globally Optimized Molecular Embeddings for Dynamic
  Reaction Solvate Shell Optimization and Active Site Design}. \emph{Top.
  Catal.} \textbf{2021}, \emph{65}, pages 281--288\relax
\mciteBstWouldAddEndPuncttrue
\mciteSetBstMidEndSepPunct{\mcitedefaultmidpunct}
{\mcitedefaultendpunct}{\mcitedefaultseppunct}\relax
\EndOfBibitem
\bibitem[Wang \latin{et~al.}(2006)Wang, Hu, Beratan, and Yang]{Wang2006}
Wang,~M.; Hu,~X.; Beratan,~D.~N.; Yang,~W. {Designing Molecules by Optimizing
  Potentials}. \emph{J. Am. Chem. Soc.} \textbf{2006}, \emph{128},
  3228--3232\relax
\mciteBstWouldAddEndPuncttrue
\mciteSetBstMidEndSepPunct{\mcitedefaultmidpunct}
{\mcitedefaultendpunct}{\mcitedefaultseppunct}\relax
\EndOfBibitem
\bibitem[Shiraogawa and Ehara(2020)Shiraogawa, and Ehara]{Shiraogawa2020}
Shiraogawa,~T.; Ehara,~M. {Theoretical Design of Photofunctional Molecular
  Aggregates for Optical Properties: An Inverse Design Approach}. \emph{J.
  Phys. Chem. C} \textbf{2020}, \emph{124}, 13329--13337\relax
\mciteBstWouldAddEndPuncttrue
\mciteSetBstMidEndSepPunct{\mcitedefaultmidpunct}
{\mcitedefaultendpunct}{\mcitedefaultseppunct}\relax
\EndOfBibitem
\bibitem[{von Lilienfeld} \latin{et~al.}(2005){von Lilienfeld}, Lins, and
  Rothlisberger]{vonLilienfeld2005}
{von Lilienfeld},~O.~A.; Lins,~R.~D.; Rothlisberger,~U. {Variational Particle
  Number Approach for Rational Compound Design}. \emph{Phys. Rev. Lett.}
  \textbf{2005}, \emph{95}, 153002\relax
\mciteBstWouldAddEndPuncttrue
\mciteSetBstMidEndSepPunct{\mcitedefaultmidpunct}
{\mcitedefaultendpunct}{\mcitedefaultseppunct}\relax
\EndOfBibitem
\bibitem[Straatsma and McCammon(1992)Straatsma, and McCammon]{Straatsma1992}
Straatsma,~T.~P.; McCammon,~J.~A. {Computational Alchemy}. \emph{Annu. Rev.
  Phys. Chem.} \textbf{1992}, \emph{43}, 407--435\relax
\mciteBstWouldAddEndPuncttrue
\mciteSetBstMidEndSepPunct{\mcitedefaultmidpunct}
{\mcitedefaultendpunct}{\mcitedefaultseppunct}\relax
\EndOfBibitem
\bibitem[Kirkwood(1935)]{Kirkwood1935}
Kirkwood,~J.~G. {Statistical Mechanics of Fluid Mixtures}. \emph{J. Chem.
  Phys.} \textbf{1935}, \emph{3}, 300--313\relax
\mciteBstWouldAddEndPuncttrue
\mciteSetBstMidEndSepPunct{\mcitedefaultmidpunct}
{\mcitedefaultendpunct}{\mcitedefaultseppunct}\relax
\EndOfBibitem
\bibitem[Mu{\~n}oz and C{\'a}rdenas(2017)Mu{\~n}oz, and
  C{\'a}rdenas]{Munoz2017}
Mu{\~n}oz,~M.; C{\'a}rdenas,~C. {How predictive could alchemical derivatives
  be?} \emph{Phys. Chem. Chem. Phys.} \textbf{2017}, \emph{19},
  16003--16012\relax
\mciteBstWouldAddEndPuncttrue
\mciteSetBstMidEndSepPunct{\mcitedefaultmidpunct}
{\mcitedefaultendpunct}{\mcitedefaultseppunct}\relax
\EndOfBibitem
\bibitem[Chang and {von Lilienfeld}(2014)Chang, and {von
  Lilienfeld}]{Chang2014}
Chang,~K. Y.~S.; {von Lilienfeld},~O.~A. {Quantum Mechanical Treatment of
  Variable Molecular Composition: From 'Alchemical' Changes of State Functions
  to Rational Compound Design}. \emph{Chimia} \textbf{2014}, \emph{68},
  602--608\relax
\mciteBstWouldAddEndPuncttrue
\mciteSetBstMidEndSepPunct{\mcitedefaultmidpunct}
{\mcitedefaultendpunct}{\mcitedefaultseppunct}\relax
\EndOfBibitem
\bibitem[{to Baben} \latin{et~al.}(2016){to Baben}, Achenbach, and {von
  Lilienfeld}]{toBaben2016}
{to Baben},~M.; Achenbach,~J.~O.; {von Lilienfeld},~O.~A. {Guiding {\emph{ab
  initio}} calculations by alchemical derivatives}. \emph{J. Chem. Phys.}
  \textbf{2016}, \emph{144}, 104103\relax
\mciteBstWouldAddEndPuncttrue
\mciteSetBstMidEndSepPunct{\mcitedefaultmidpunct}
{\mcitedefaultendpunct}{\mcitedefaultseppunct}\relax
\EndOfBibitem
\bibitem[Saravanan \latin{et~al.}(2017)Saravanan, Kitchin, {von Lilienfeld},
  and Keith]{Saravanan2017}
Saravanan,~K.; Kitchin,~J.~R.; {von Lilienfeld},~O.~A.; Keith,~J.~A.
  {Alchemical Predictions for Computational Catalysis: Potential and
  Limitations}. \emph{J. Phys. Chem. Lett.} \textbf{2017}, \emph{8},
  5002--5007\relax
\mciteBstWouldAddEndPuncttrue
\mciteSetBstMidEndSepPunct{\mcitedefaultmidpunct}
{\mcitedefaultendpunct}{\mcitedefaultseppunct}\relax
\EndOfBibitem
\bibitem[Domenichini \latin{et~al.}(2020)Domenichini, {von Rudorff}, and {von
  Lilienfeld}]{Domenichini2020}
Domenichini,~G.; {von Rudorff},~G.~F.; {von Lilienfeld},~O.~A. {Effects of
  perturbation order and basis set on alchemical predictions}. \emph{J. Chem.
  Phys.} \textbf{2020}, \emph{153}, 144118\relax
\mciteBstWouldAddEndPuncttrue
\mciteSetBstMidEndSepPunct{\mcitedefaultmidpunct}
{\mcitedefaultendpunct}{\mcitedefaultseppunct}\relax
\EndOfBibitem
\bibitem[{Sanchez-Lengeling} and {Aspuru-Guzik}(2018){Sanchez-Lengeling}, and
  {Aspuru-Guzik}]{Sanchez-lengeling2018}
{Sanchez-Lengeling},~B.; {Aspuru-Guzik},~A. {Inverse molecular design using
  machine learning: Generative models for matter engineering}. \emph{Science}
  \textbf{2018}, \emph{361}, 360--365\relax
\mciteBstWouldAddEndPuncttrue
\mciteSetBstMidEndSepPunct{\mcitedefaultmidpunct}
{\mcitedefaultendpunct}{\mcitedefaultseppunct}\relax
\EndOfBibitem
\bibitem[Mater and Coote(2019)Mater, and Coote]{Mater2019}
Mater,~A.~C.; Coote,~M.~L. {Deep Learning in Chemistry}. \emph{J. Chem. Inf.
  Model.} \textbf{2019}, \emph{59}, 2545--2559\relax
\mciteBstWouldAddEndPuncttrue
\mciteSetBstMidEndSepPunct{\mcitedefaultmidpunct}
{\mcitedefaultendpunct}{\mcitedefaultseppunct}\relax
\EndOfBibitem
\bibitem[{von Lilienfeld} \latin{et~al.}(2020){von Lilienfeld}, M{\"u}ller, and
  Tkatchenko]{vonLilienfeld2020}
{von Lilienfeld},~O.~A.; M{\"u}ller,~K.-R.; Tkatchenko,~A. {Exploring chemical
  compound space with quantum-based machine learning}. \emph{Nat. Rev. Chem.}
  \textbf{2020}, \emph{4}, 347--358\relax
\mciteBstWouldAddEndPuncttrue
\mciteSetBstMidEndSepPunct{\mcitedefaultmidpunct}
{\mcitedefaultendpunct}{\mcitedefaultseppunct}\relax
\EndOfBibitem
\bibitem[Jena and Sun(2021)Jena, and Sun]{Jena2021}
Jena,~P.; Sun,~Q. {Theory-Guided Discovery of Novel Materials}. \emph{J. Phys.
  Chem. Lett.} \textbf{2021}, \emph{12}, 6499--6513\relax
\mciteBstWouldAddEndPuncttrue
\mciteSetBstMidEndSepPunct{\mcitedefaultmidpunct}
{\mcitedefaultendpunct}{\mcitedefaultseppunct}\relax
\EndOfBibitem
\bibitem[Jena and Sun(2021)Jena, and Sun]{Jena2021a}
Jena,~P.; Sun,~Q. {Correction to ``Theory-Guided Discovery of Novel
  Materials''}. \emph{J. Phys. Chem. Lett.} \textbf{2021}, \emph{12},
  7490\relax
\mciteBstWouldAddEndPuncttrue
\mciteSetBstMidEndSepPunct{\mcitedefaultmidpunct}
{\mcitedefaultendpunct}{\mcitedefaultseppunct}\relax
\EndOfBibitem
\bibitem[Nandy \latin{et~al.}(2021)Nandy, Duan, Taylor, Liu, Steeves, and
  Kulik]{Nandy2021}
Nandy,~A.; Duan,~C.; Taylor,~M.~G.; Liu,~F.; Steeves,~A.~H.; Kulik,~H.~J.
  {Computational Discovery of Transition-metal Complexes: From High-throughput
  Screening to Machine Learning}. \emph{Chem. Rev.} \textbf{2021}, \emph{121},
  9927--10000\relax
\mciteBstWouldAddEndPuncttrue
\mciteSetBstMidEndSepPunct{\mcitedefaultmidpunct}
{\mcitedefaultendpunct}{\mcitedefaultseppunct}\relax
\EndOfBibitem
\bibitem[Janet \latin{et~al.}(2021)Janet, Duan, Nandy, Liu, and
  Kulik]{Janet2021}
Janet,~J.~P.; Duan,~C.; Nandy,~A.; Liu,~F.; Kulik,~H.~J. {Navigating
  Transition-Metal Chemical Space: Artificial Intelligence for First-Principles
  Design}. \emph{Acc. Chem. Res.} \textbf{2021}, \emph{54}, 532--545\relax
\mciteBstWouldAddEndPuncttrue
\mciteSetBstMidEndSepPunct{\mcitedefaultmidpunct}
{\mcitedefaultendpunct}{\mcitedefaultseppunct}\relax
\EndOfBibitem
\bibitem[Huang and {von Lilienfeld}(2021)Huang, and {von
  Lilienfeld}]{Huang2021}
Huang,~B.; {von Lilienfeld},~O.~A. {Ab Initio Machine Learning in Chemical
  Compound Space}. \emph{Chem. Rev.} \textbf{2021}, \emph{121},
  1000--10036\relax
\mciteBstWouldAddEndPuncttrue
\mciteSetBstMidEndSepPunct{\mcitedefaultmidpunct}
{\mcitedefaultendpunct}{\mcitedefaultseppunct}\relax
\EndOfBibitem
\bibitem[Teunissen \latin{et~al.}(2019)Teunissen, De~Proft, and
  De~Vleeschouwer]{Teunissen2019}
Teunissen,~J.~L.; De~Proft,~F.; De~Vleeschouwer,~F. {Acceleration of Inverse
  Molecular Design by Using Predictive Techniques}. \emph{J. Chem. Inf. Model.}
  \textbf{2019}, \emph{59}, 2587--2599\relax
\mciteBstWouldAddEndPuncttrue
\mciteSetBstMidEndSepPunct{\mcitedefaultmidpunct}
{\mcitedefaultendpunct}{\mcitedefaultseppunct}\relax
\EndOfBibitem
\bibitem[Patra \latin{et~al.}(2017)Patra, Meenakshisundaram, Hung, and
  Simmons]{Patra2017}
Patra,~T.~K.; Meenakshisundaram,~V.; Hung,~J.-H.; Simmons,~D.~S.
  {Neural-Network-Biased Genetic Algorithms for Materials Design: Evolutionary
  Algorithms That Learn}. \emph{ACS Comb. Sci.} \textbf{2017}, \emph{19},
  96--107\relax
\mciteBstWouldAddEndPuncttrue
\mciteSetBstMidEndSepPunct{\mcitedefaultmidpunct}
{\mcitedefaultendpunct}{\mcitedefaultseppunct}\relax
\EndOfBibitem
\bibitem[{Mannodi-Kanakkithodi} \latin{et~al.}(2016){Mannodi-Kanakkithodi},
  Pilania, Huan, Lookman, and Ramprasad]{Mannodi-Kanakkithodi2016}
{Mannodi-Kanakkithodi},~A.; Pilania,~G.; Huan,~T.~D.; Lookman,~T.;
  Ramprasad,~R. {Machine Learning Strategy for Accelerated Design of Polymer
  Dielectrics}. \emph{Sci. Rep.} \textbf{2016}, \emph{6}, 20952\relax
\mciteBstWouldAddEndPuncttrue
\mciteSetBstMidEndSepPunct{\mcitedefaultmidpunct}
{\mcitedefaultendpunct}{\mcitedefaultseppunct}\relax
\EndOfBibitem
\bibitem[Hautier \latin{et~al.}(2010)Hautier, Fischer, Jain, Mueller, and
  Ceder]{Hautier2010}
Hautier,~G.; Fischer,~C.~C.; Jain,~A.; Mueller,~T.; Ceder,~G. {Finding Nature's
  Missing Ternary Oxide Compounds Using Machine Learning and Density Functional
  Theory}. \emph{Chem. Mater.} \textbf{2010}, \emph{22}, 3762--3767\relax
\mciteBstWouldAddEndPuncttrue
\mciteSetBstMidEndSepPunct{\mcitedefaultmidpunct}
{\mcitedefaultendpunct}{\mcitedefaultseppunct}\relax
\EndOfBibitem
\bibitem[Raccuglia \latin{et~al.}(2016)Raccuglia, Elbert, Adler, Falk, Wenny,
  Mollo, Zeller, Friedler, Schrier, and Norquist]{Raccuglia2016}
Raccuglia,~P.; Elbert,~K.~C.; Adler,~P. D.~F.; Falk,~C.; Wenny,~M.~B.;
  Mollo,~A.; Zeller,~M.; Friedler,~S.~A.; Schrier,~J.; Norquist,~A.~J.
  {Machine-learning-assisted materials discovery using failed experiments}.
  \emph{Nature} \textbf{2016}, \emph{533}, 73--76\relax
\mciteBstWouldAddEndPuncttrue
\mciteSetBstMidEndSepPunct{\mcitedefaultmidpunct}
{\mcitedefaultendpunct}{\mcitedefaultseppunct}\relax
\EndOfBibitem
\bibitem[Zhou \latin{et~al.}(2017)Zhou, Li, and Zare]{Zhou2017}
Zhou,~Z.; Li,~X.; Zare,~R.~N. {Optimizing Chemical Reactions with Deep
  Reinforcement Learning}. \emph{ACS Cent. Sci.} \textbf{2017}, \emph{3},
  1337--1344\relax
\mciteBstWouldAddEndPuncttrue
\mciteSetBstMidEndSepPunct{\mcitedefaultmidpunct}
{\mcitedefaultendpunct}{\mcitedefaultseppunct}\relax
\EndOfBibitem
\bibitem[Janet \latin{et~al.}(2020)Janet, Ramesh, Duan, and Kulik]{Janet2020}
Janet,~J.~P.; Ramesh,~S.; Duan,~C.; Kulik,~H.~J. {Accurate Multiobjective
  Design in a Space of Millions of Transition Metal Complexes with
  Neural-Network-Driven Efficient Global Optimization}. \emph{ACS Cent. Sci.}
  \textbf{2020}, \emph{6}, 513--524\relax
\mciteBstWouldAddEndPuncttrue
\mciteSetBstMidEndSepPunct{\mcitedefaultmidpunct}
{\mcitedefaultendpunct}{\mcitedefaultseppunct}\relax
\EndOfBibitem
\bibitem[Proppe \latin{et~al.}(2016)Proppe, Husch, Simm, and
  Reiher]{Proppe2016}
Proppe,~J.; Husch,~T.; Simm,~G.~N.; Reiher,~M. {Uncertainty quantification for
  quantum chemical models of complex reaction networks}. \emph{Faraday
  Discuss.} \textbf{2016}, \emph{195}, 497--520\relax
\mciteBstWouldAddEndPuncttrue
\mciteSetBstMidEndSepPunct{\mcitedefaultmidpunct}
{\mcitedefaultendpunct}{\mcitedefaultseppunct}\relax
\EndOfBibitem
\bibitem[Proppe and Reiher(2019)Proppe, and Reiher]{Proppe2019}
Proppe,~J.; Reiher,~M. {Mechanism Deduction from Noisy Chemical Reaction
  Networks}. \emph{J. Chem. Theory Comput.} \textbf{2019}, \emph{15},
  357--370\relax
\mciteBstWouldAddEndPuncttrue
\mciteSetBstMidEndSepPunct{\mcitedefaultmidpunct}
{\mcitedefaultendpunct}{\mcitedefaultseppunct}\relax
\EndOfBibitem
\bibitem[Reiher(2021)]{Reiher2021}
Reiher,~M. {Molecule-specific Uncertainty Quantification in Quantum Chemical
  Studies}. \emph{Isr. J. Chem.} \textbf{2021}, \emph{62}, e202100101\relax
\mciteBstWouldAddEndPuncttrue
\mciteSetBstMidEndSepPunct{\mcitedefaultmidpunct}
{\mcitedefaultendpunct}{\mcitedefaultseppunct}\relax
\EndOfBibitem
\bibitem[iso(2008)]{iso_iec_guide_98_3}
{ISO/IEC Guide 98-3:2008 Uncertainty of measurement --- Part 3: Guide to the
  expression of uncertainty in measurement (GUM:1995)}. 2008\relax
\mciteBstWouldAddEndPuncttrue
\mciteSetBstMidEndSepPunct{\mcitedefaultmidpunct}
{\mcitedefaultendpunct}{\mcitedefaultseppunct}\relax
\EndOfBibitem
\bibitem[Irikura \latin{et~al.}(2004)Irikura, Johnson, and Kacker]{Irikura2004}
Irikura,~K.~K.; Johnson,~R.~D.,~III; Kacker,~R.~N. {Uncertainty associated with
  virtual measurements from computational quantum chemistry models}.
  \emph{Metrologia} \textbf{2004}, \emph{41}, 369--375\relax
\mciteBstWouldAddEndPuncttrue
\mciteSetBstMidEndSepPunct{\mcitedefaultmidpunct}
{\mcitedefaultendpunct}{\mcitedefaultseppunct}\relax
\EndOfBibitem
\bibitem[Pernot \latin{et~al.}(2015)Pernot, Civalleri, Presti, and
  Savin]{Pernot2015}
Pernot,~P.; Civalleri,~B.; Presti,~D.; Savin,~A. {Prediction Uncertainty of
  Density Functional Approximations for Properties of Crystals with Cubic
  Symmetry}. \emph{J. Phys. Chem. A} \textbf{2015}, \emph{119},
  5288--5304\relax
\mciteBstWouldAddEndPuncttrue
\mciteSetBstMidEndSepPunct{\mcitedefaultmidpunct}
{\mcitedefaultendpunct}{\mcitedefaultseppunct}\relax
\EndOfBibitem
\bibitem[Scott and Radom(1996)Scott, and Radom]{Scott1996}
Scott,~A.~P.; Radom,~L. {Harmonic Vibrational Frequencies: An Evaluation of
  Hartree--Fock, M\o ller--Plesset, Quadratic Configuration Interaction,
  Density Functional Theory, and Semiempirical Scale Factors}. \emph{J. Phys.
  Chem.} \textbf{1996}, \emph{100}, 16502--16513\relax
\mciteBstWouldAddEndPuncttrue
\mciteSetBstMidEndSepPunct{\mcitedefaultmidpunct}
{\mcitedefaultendpunct}{\mcitedefaultseppunct}\relax
\EndOfBibitem
\bibitem[Irikura \latin{et~al.}(2005)Irikura, Johnson, and Kacker]{Irikura2005}
Irikura,~K.~K.; Johnson,~R.~D.,~III; Kacker,~R.~N. {Uncertainties in Scaling
  Factors for ab Initio Vibrational Frequencies}. \emph{J. Phys. Chem. A}
  \textbf{2005}, \emph{109}, 8430--8437\relax
\mciteBstWouldAddEndPuncttrue
\mciteSetBstMidEndSepPunct{\mcitedefaultmidpunct}
{\mcitedefaultendpunct}{\mcitedefaultseppunct}\relax
\EndOfBibitem
\bibitem[Pernot and Cailliez(2017)Pernot, and Cailliez]{Pernot2017}
Pernot,~P.; Cailliez,~F. {A Critical Review of Statistical
  Calibration/Prediction Models Handling Data Inconsistency and Model
  Inadequacy}. \emph{AIChE J.} \textbf{2017}, \emph{63}, 4642--4665\relax
\mciteBstWouldAddEndPuncttrue
\mciteSetBstMidEndSepPunct{\mcitedefaultmidpunct}
{\mcitedefaultendpunct}{\mcitedefaultseppunct}\relax
\EndOfBibitem
\bibitem[Simm \latin{et~al.}(2017)Simm, Proppe, and Reiher]{Simm2017}
Simm,~G.~N.; Proppe,~J.; Reiher,~M. {Error Assessment of Computational Models
  in Chemistry}. \emph{Chimia} \textbf{2017}, \emph{71}, 202--208\relax
\mciteBstWouldAddEndPuncttrue
\mciteSetBstMidEndSepPunct{\mcitedefaultmidpunct}
{\mcitedefaultendpunct}{\mcitedefaultseppunct}\relax
\EndOfBibitem
\bibitem[Mortensen \latin{et~al.}(2005)Mortensen, Kaasbjerg, Frederiksen,
  N{\o}rskov, Sethna, and Jacobsen]{Mortensen2005}
Mortensen,~J.~J.; Kaasbjerg,~K.; Frederiksen,~S.~L.; N{\o}rskov,~J.~K.;
  Sethna,~J.~P.; Jacobsen,~K.~W. {Bayesian Error Estimation in
  Density-Functional Theory}. \emph{Phys. Rev. Lett.} \textbf{2005}, \emph{95},
  216401\relax
\mciteBstWouldAddEndPuncttrue
\mciteSetBstMidEndSepPunct{\mcitedefaultmidpunct}
{\mcitedefaultendpunct}{\mcitedefaultseppunct}\relax
\EndOfBibitem
\bibitem[Aldegunde \latin{et~al.}(2016)Aldegunde, Kermode, and
  Zabaras]{Aldegunde2016}
Aldegunde,~M.; Kermode,~J.~R.; Zabaras,~N. {Development of an
  exchange--correlation functional with uncertainty quantification capabilities
  for density functional theory}. \emph{J. Comput. Phys.} \textbf{2016},
  \emph{311}, 173--195\relax
\mciteBstWouldAddEndPuncttrue
\mciteSetBstMidEndSepPunct{\mcitedefaultmidpunct}
{\mcitedefaultendpunct}{\mcitedefaultseppunct}\relax
\EndOfBibitem
\bibitem[Simm and Reiher(2016)Simm, and Reiher]{Simm2016}
Simm,~G.~N.; Reiher,~M. {Systematic Error Estimation for Chemical Reaction
  Energies}. \emph{J. Chem. Theory Comput.} \textbf{2016}, \emph{12},
  2762--2773\relax
\mciteBstWouldAddEndPuncttrue
\mciteSetBstMidEndSepPunct{\mcitedefaultmidpunct}
{\mcitedefaultendpunct}{\mcitedefaultseppunct}\relax
\EndOfBibitem
\bibitem[Pernot(2017)]{Pernot2017a}
Pernot,~P. {The parameter uncertainty inflation fallacy}. \emph{J. Chem. Phys.}
  \textbf{2017}, \emph{147}, 104102\relax
\mciteBstWouldAddEndPuncttrue
\mciteSetBstMidEndSepPunct{\mcitedefaultmidpunct}
{\mcitedefaultendpunct}{\mcitedefaultseppunct}\relax
\EndOfBibitem
\bibitem[Efron(1979)]{Efron1979}
Efron,~B. {Bootstrap Methods: Another Look at the Jackknife}. \emph{Ann. Stat.}
  \textbf{1979}, \emph{7}, 1--26\relax
\mciteBstWouldAddEndPuncttrue
\mciteSetBstMidEndSepPunct{\mcitedefaultmidpunct}
{\mcitedefaultendpunct}{\mcitedefaultseppunct}\relax
\EndOfBibitem
\bibitem[Proppe and Reiher(2017)Proppe, and Reiher]{Proppe2017}
Proppe,~J.; Reiher,~M. {Reliable Estimation of Prediction Uncertainty for
  Physicochemical Property Models}. \emph{J. Chem. Theory Comput.}
  \textbf{2017}, \emph{13}, 3297--3317\relax
\mciteBstWouldAddEndPuncttrue
\mciteSetBstMidEndSepPunct{\mcitedefaultmidpunct}
{\mcitedefaultendpunct}{\mcitedefaultseppunct}\relax
\EndOfBibitem
\bibitem[Weymuth \latin{et~al.}(2018)Weymuth, Proppe, and Reiher]{Weymuth2018}
Weymuth,~T.; Proppe,~J.; Reiher,~M. {Statistical Analysis of Semiclassical
  Dispersion Corrections}. \emph{J. Chem. Theory Comput.} \textbf{2018},
  \emph{14}, 2480--2494\relax
\mciteBstWouldAddEndPuncttrue
\mciteSetBstMidEndSepPunct{\mcitedefaultmidpunct}
{\mcitedefaultendpunct}{\mcitedefaultseppunct}\relax
\EndOfBibitem
\bibitem[Frenklach \latin{et~al.}(2002)Frenklach, Packard, and
  Seiler]{Frenklach2002}
Frenklach,~M.; Packard,~A.; Seiler,~P. {Prediction uncertainty from models and
  data}. Proceedings of the 2002 American Control Conference. 2002; pp
  4135--4140\relax
\mciteBstWouldAddEndPuncttrue
\mciteSetBstMidEndSepPunct{\mcitedefaultmidpunct}
{\mcitedefaultendpunct}{\mcitedefaultseppunct}\relax
\EndOfBibitem
\bibitem[Frenklach \latin{et~al.}(2004)Frenklach, Packard, Seiler, and
  Feeley]{Frenklach2004}
Frenklach,~M.; Packard,~A.; Seiler,~P.; Feeley,~R. {Collaborative Data
  Processing in Developing Predictive Models of Complex Reaction Systems}.
  \emph{Int. J. Chem. Kinet.} \textbf{2004}, \emph{36}, 57--66\relax
\mciteBstWouldAddEndPuncttrue
\mciteSetBstMidEndSepPunct{\mcitedefaultmidpunct}
{\mcitedefaultendpunct}{\mcitedefaultseppunct}\relax
\EndOfBibitem
\bibitem[Russi \latin{et~al.}(2010)Russi, Packard, and Frenklach]{Russi2010}
Russi,~T.; Packard,~A.; Frenklach,~M. {Uncertainty quantification: Making
  predictions of complex reaction systems reliable}. \emph{Chem. Phys. Lett.}
  \textbf{2010}, \emph{499}, 1--8\relax
\mciteBstWouldAddEndPuncttrue
\mciteSetBstMidEndSepPunct{\mcitedefaultmidpunct}
{\mcitedefaultendpunct}{\mcitedefaultseppunct}\relax
\EndOfBibitem
\bibitem[Frenklach \latin{et~al.}(1992)Frenklach, Wang, and
  Rabinowitz]{Frenklach1992}
Frenklach,~M.; Wang,~H.; Rabinowitz,~M.~J. {Optimization and analysis of large
  chemical kinetic mechanisms using the solution mapping method---combustion of
  methane}. \emph{Progr. Energy Combust. Sci.} \textbf{1992}, \emph{18},
  47--73\relax
\mciteBstWouldAddEndPuncttrue
\mciteSetBstMidEndSepPunct{\mcitedefaultmidpunct}
{\mcitedefaultendpunct}{\mcitedefaultseppunct}\relax
\EndOfBibitem
\bibitem[Oreluk \latin{et~al.}(2018)Oreluk, Liu, Hegde, Li, Packard, Frenklach,
  and Zubarev]{Oreluk2018}
Oreluk,~J.; Liu,~Z.; Hegde,~A.; Li,~W.; Packard,~A.; Frenklach,~M.; Zubarev,~D.
  {Diagnostics of Data-Driven Models: Uncertainty Quantification of PM7
  Semi-Empirical Quantum Chemical Method}. \emph{Sci. Rep.} \textbf{2018},
  \emph{8}, 13248\relax
\mciteBstWouldAddEndPuncttrue
\mciteSetBstMidEndSepPunct{\mcitedefaultmidpunct}
{\mcitedefaultendpunct}{\mcitedefaultseppunct}\relax
\EndOfBibitem
\bibitem[Oung \latin{et~al.}(2018)Oung, Rudolph, and Jacob]{Oung2018}
Oung,~S.~W.; Rudolph,~J.; Jacob,~C.~R. {Uncertainty quantification in
  theoretical spectroscopy: The structural sensitivity of X-ray emission
  spectra}. \emph{Int. J. Quantum Chem.} \textbf{2018}, \emph{118},
  e25458\relax
\mciteBstWouldAddEndPuncttrue
\mciteSetBstMidEndSepPunct{\mcitedefaultmidpunct}
{\mcitedefaultendpunct}{\mcitedefaultseppunct}\relax
\EndOfBibitem
\bibitem[Bergmann \latin{et~al.}(2020)Bergmann, Welzel, and
  Jacob]{Bergmann2020}
Bergmann,~T.~G.; Welzel,~M.~O.; Jacob,~C.~R. {Towards theoretical spectroscopy
  with error bars: systematic quantification of the structural sensitivity of
  calculated spectra}. \emph{Chem. Sci.} \textbf{2020}, \emph{11},
  1862--1877\relax
\mciteBstWouldAddEndPuncttrue
\mciteSetBstMidEndSepPunct{\mcitedefaultmidpunct}
{\mcitedefaultendpunct}{\mcitedefaultseppunct}\relax
\EndOfBibitem
\bibitem[Simm and Reiher(2018)Simm, and Reiher]{Simm2018}
Simm,~G.~N.; Reiher,~M. {Error-Controlled Exploration of Chemical Reaction
  Networks with Gaussian Processes}. \emph{J. Chem. Theory Comput.}
  \textbf{2018}, \emph{14}, 5238--5248\relax
\mciteBstWouldAddEndPuncttrue
\mciteSetBstMidEndSepPunct{\mcitedefaultmidpunct}
{\mcitedefaultendpunct}{\mcitedefaultseppunct}\relax
\EndOfBibitem
\bibitem[Rasmussen(2004)]{Rasmussen2004}
Rasmussen,~C.~E. In \emph{{Advanced Lectures on Machine Learning}};
  Bousquet,~O., {von Luxburg},~U., R{\"a}tsch,~G., Eds.; Lecture Notes in
  Computer Science; Springer: Berlin, Germany, 2004; Vol. 3176; pp 63--71\relax
\mciteBstWouldAddEndPuncttrue
\mciteSetBstMidEndSepPunct{\mcitedefaultmidpunct}
{\mcitedefaultendpunct}{\mcitedefaultseppunct}\relax
\EndOfBibitem
\bibitem[Peterson \latin{et~al.}(2017)Peterson, Christensen, and
  Khorshidi]{Peterson2017}
Peterson,~A.~A.; Christensen,~R.; Khorshidi,~A. {Addressing uncertainty in
  atomistic machine learning}. \emph{Phys. Chem. Chem. Phys.} \textbf{2017},
  \emph{19}, 10978--10985\relax
\mciteBstWouldAddEndPuncttrue
\mciteSetBstMidEndSepPunct{\mcitedefaultmidpunct}
{\mcitedefaultendpunct}{\mcitedefaultseppunct}\relax
\EndOfBibitem
\bibitem[Musil \latin{et~al.}(2019)Musil, Willatt, Langovoy, and
  Ceriotti]{Musil2019}
Musil,~F.; Willatt,~M.~J.; Langovoy,~M.~A.; Ceriotti,~M. {Fast and Accurate
  Uncertainty Estimation in Chemical Machine Learning}. \emph{J. Chem. Theory
  Comput.} \textbf{2019}, \emph{15}, 906--915\relax
\mciteBstWouldAddEndPuncttrue
\mciteSetBstMidEndSepPunct{\mcitedefaultmidpunct}
{\mcitedefaultendpunct}{\mcitedefaultseppunct}\relax
\EndOfBibitem
\bibitem[Vishwakarma \latin{et~al.}(2021)Vishwakarma, Sonpal, and
  Hachmann]{Vishwakarma2021}
Vishwakarma,~G.; Sonpal,~A.; Hachmann,~J. {Metrics for Benchmarking and
  Uncertainty Quantification: Quality, Applicability, and Best Practices for
  Machine Learning in Chemistry}. \emph{Trends Chem.} \textbf{2021}, \emph{3},
  146--156\relax
\mciteBstWouldAddEndPuncttrue
\mciteSetBstMidEndSepPunct{\mcitedefaultmidpunct}
{\mcitedefaultendpunct}{\mcitedefaultseppunct}\relax
\EndOfBibitem
\bibitem[Venturi \latin{et~al.}(2020)Venturi, Jaffe, and Panesi]{Venturi2020}
Venturi,~S.; Jaffe,~R.~L.; Panesi,~M. {Bayesian Machine Learning Approach to
  the Quantification of Uncertainties on Ab Initio Potential Energy Surfaces}.
  \emph{J. Phys. Chem. A} \textbf{2020}, \emph{124}, 5129--5146\relax
\mciteBstWouldAddEndPuncttrue
\mciteSetBstMidEndSepPunct{\mcitedefaultmidpunct}
{\mcitedefaultendpunct}{\mcitedefaultseppunct}\relax
\EndOfBibitem
\bibitem[Liu and Wallqvist(2019)Liu, and Wallqvist]{Liu2019}
Liu,~R.; Wallqvist,~A. {Molecular Similarity-Based Domain Applicability Metric
  Efficiently Identifies Out-of-Domain Compounds}. \emph{J. Chem. Inf. Model.}
  \textbf{2019}, \emph{59}, 181--189\relax
\mciteBstWouldAddEndPuncttrue
\mciteSetBstMidEndSepPunct{\mcitedefaultmidpunct}
{\mcitedefaultendpunct}{\mcitedefaultseppunct}\relax
\EndOfBibitem
\bibitem[Liu \latin{et~al.}(2018)Liu, Glover, Feasel, and Wallqvist]{Liu2018}
Liu,~R.; Glover,~K.~P.; Feasel,~M.~G.; Wallqvist,~A. {General Approach to
  Estimate Error Bars for Quantitative Structure--Activity Relationship
  Predictions of Molecular Activity}. \emph{J. Chem. Inf. Model.}
  \textbf{2018}, \emph{58}, 1561--1575\relax
\mciteBstWouldAddEndPuncttrue
\mciteSetBstMidEndSepPunct{\mcitedefaultmidpunct}
{\mcitedefaultendpunct}{\mcitedefaultseppunct}\relax
\EndOfBibitem
\bibitem[Janet \latin{et~al.}(2019)Janet, Duan, Yang, Nandy, and
  Kulik]{Janet2019}
Janet,~J.~P.; Duan,~C.; Yang,~T.; Nandy,~A.; Kulik,~H.~J. {A quantitative
  uncertainty metric controls error in neural network-driven chemical
  discovery}. \emph{Chem. Sci.} \textbf{2019}, \emph{10}, 7913--7922\relax
\mciteBstWouldAddEndPuncttrue
\mciteSetBstMidEndSepPunct{\mcitedefaultmidpunct}
{\mcitedefaultendpunct}{\mcitedefaultseppunct}\relax
\EndOfBibitem
\bibitem[Brown \latin{et~al.}(2019)Brown, Fiscato, Segler, and
  Vaucher]{Brown2019}
Brown,~N.; Fiscato,~M.; Segler,~M.~H.; Vaucher,~A.~C. {GuacaMol: Benchmarking
  Models for de Novo Molecular Design}. \emph{J. Chem. Inf. Model.}
  \textbf{2019}, \emph{59}, 1096--1108\relax
\mciteBstWouldAddEndPuncttrue
\mciteSetBstMidEndSepPunct{\mcitedefaultmidpunct}
{\mcitedefaultendpunct}{\mcitedefaultseppunct}\relax
\EndOfBibitem
\end{mcitethebibliography}
\providecommand{\latin}[1]{#1}
\makeatletter
\providecommand{\doi}
  {\begingroup\let\do\@makeother\dospecials
  \catcode`\{=1 \catcode`\}=2 \doi@aux}
\providecommand{\doi@aux}[1]{\endgroup\texttt{#1}}
\makeatother
\providecommand*\mcitethebibliography{\thebibliography}
\csname @ifundefined\endcsname{endmcitethebibliography}
  {\let\endmcitethebibliography\endthebibliography}{}

\end{document}